\documentclass[aps,pra,11pt,amsmath,notitlepage,tightenlines,superscriptaddress,floatfix,eqsecnum,showpacs,longbibliography]{revtex4-1}
\usepackage{amssymb,amsmath}
\usepackage{accents}
\usepackage{subdepth}
\usepackage{graphicx}
\usepackage{sansmath}

\usepackage{xcolor}
\definecolor{dark-red}{rgb}{0.4,0.15,0.15}
\definecolor{dark-blue}{rgb}{0.15,0.15,0.4}
\definecolor{medium-blue}{rgb}{0,0,0.5}
\usepackage{hyperref}
\usepackage[all]{hypcap}
\usepackage{cleveref}
\hypersetup{
    colorlinks, linkcolor={dark-red},
    citecolor={dark-blue}, urlcolor={medium-blue}
} 

\newcommand{\nsp}{\hspace{-0.4pt}}
\newcommand{\ssp}{\hspace{0.4pt}}

\newcommand{\ket}[1]{\lvert\ssp #1\ssp \rangle}
\newcommand{\bra}[1]{\langle\ssp #1\ssp \rvert}
\newcommand{\bigket}[1]{\big\lvert\ssp #1\ssp\big\rangle}
\newcommand{\bigbra}[1]{\big\langle\ssp#1\ssp\big\rvert}
\newcommand{\norm}[1]{\lvert #1 \rvert}

\newcommand{\commutb}[2]{\big[\ssp #1\ssp,\,#2\ssp\big]}
\newcommand{\identity}{\openone}
\newcommand{\half}{\frac{1}{2}}
\newcommand{\dt}[1]{\accentset{\vspace{0.4pt}\hspace{0.5pt}\mbox{\Large .}}{#1}} %

\newcommand{\sN}{\mathcal{N}}
\newcommand{\sR}{\mathcal{U}_\mathrm{gp}}
\newcommand{\sRperp}{\mathcal{U}_{\mathrm{gp}\perp}}

\newcommand{\sU}{\mathcal{U}}

\newcommand{\vac}{\mathrm{\bf vac}}
\newcommand{\psinot}{\phi}
\newcommand{\psinotnot}{\chi}
\newcommand{\braket}[2]{\langle\, #1\,\vert\, #2 \,\rangle}
\newcommand{\vecnorm}[1]{\braket{#1}{#1}}
\renewcommand{\a}{{a}}

\newcommand{\di}[2]{\frac{d\ssp #1}{d\ssp #2}}
\newcommand{\mathsp}[1]{\mbox{\sansmath $#1$}}
\newcommand{\sF}{\mathcal{F}}
\newcommand{\sP}{\mathcal{P}}
\newcommand{\proj}[1]{\ket{#1}\nsp\bra{#1}}
\newcommand\independent{\protect\mathpalette{\protect\independenT}{\scriptstyle\perp}}
\def\independenT#1#2{\mathrel{\rlap{$#1#2$}\mkern2mu{#1#2}}}
\newcommand{\doubleperp}{{\nsp\mbox{$\independent$}}}
\newcommand{\pa}[2]{\frac{\partial #1}{\partial #2}}
\newcommand{\COLON}{\mbox{\raisebox{-2pt}{\LARGE{\bf:}}}}

\newcommand{\dif}{d}
\newcommand{\uppsi}{{\Psi}}
\newcommand{\psibf}{{\boldsymbol\psi}}

\newcommand{\Hrem}{K}

\begin{document}

\title{Particle-Number-Conserving Bogoliubov Approximation \texorpdfstring{\\}{} for Bose-Einstein Condensates Using Extended Catalytic States}

\date{\today}

\author{Zhang Jiang}
\email{zhang.jiang@nasa.gov}
\affiliation{Center for Quantum Information and Control, University of New Mexico, MSC07-4220, Albuquerque, New Mexico 87131-0001, USA}
\affiliation{NASA Ames Research Center Quantum Artificial Intelligence Laboratory (QuAIL), Mail Stop 269-1, 94035 Moffett Field CA}
\affiliation{Stinger Ghaffarian Technologies Inc., 7701 Greenbelt Rd., Suite 400, Greenbelt, MD 20770}

\author{Carlton M. Caves}
\email{ccaves@unm.edu}
\affiliation{Center for Quantum Information and Control, University of New Mexico, MSC07-4220, Albuquerque, New Mexico 87131-0001, USA}
\affiliation{Centre for Engineered Quantum Systems, School of Mathematics and Physics,
University of Queensland, Brisbane, Queensland 4072, Australia}

\begin{abstract}
We encode the many-body wavefunction of a Bose-Einstein condensate (BEC) in the $N$-particle sector of an extended catalytic state.  This catalytic state is a coherent state for the condensate mode and an arbitrary state for the modes orthogonal to the condensate mode.  Going to a time-dependent interaction picture where the state of the condensate mode is displaced to the vacuum, we can organize the effective Hamiltonian by powers of ${N}^{-1/2}$.   Requiring the terms of order ${N}^{1/2}$ to vanish gives the Gross-Pitaevskii equation.  Going to the next order, $N^0$, we derive equations for the number-conserving Bogoliubov approximation, first given by Castin and Dum [Phys.\ Rev.~A {\bf 57}, 3008 (1998)].  In contrast to other approaches, ours is well suited to calculating the state evolution in the Schr\"{o}dinger picture; moreover, it is straightforward to generalize our method to multi-component BECs and to higher-order corrections.
\end{abstract}


\maketitle

\section{Introduction}
\label{sec:intro}

We consider the ground state and dynamics of a dilute-gas BEC of $N$ bosonic atoms trapped in an arbitrary external potential.  In order to describe how interparticle correlations modify the Gross-Pitaevskii (GP) equation, we go to the next level of approximation, the Bogoliubov approximation.  The Bogoliubov approximation~\cite{bogoliubov_theory_1947, fetter_nonuniform_1972, huang_statistical_1987} is important for several reasons: (i)~it tells when the Gross-Pitaevskii (mean-field) approach begins to break down; (ii)~it describes small deviations from the Gross-Pitaevskii equation and can be used to study the stability of a BEC; (iii)~it enables the calculation of how impurities change the behavior of a BEC; and (iv)~it is useful for studying phase coherence between BECs.

\begin{figure} 
   \centering
   \includegraphics[width=0.75\textwidth]{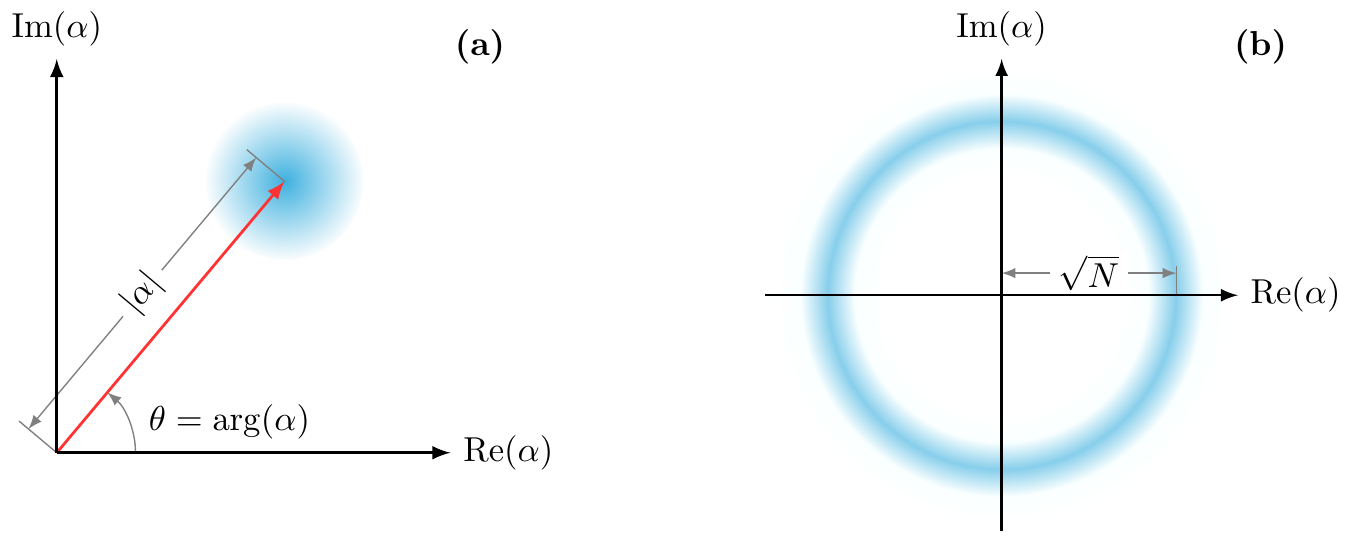}
   \caption[Phase-space representations for coherent and number states]{Phase-space representations for (a)~a coherent state with complex amplitude $\alpha$ and (b)~a number state with particle number $N$.  A number state is distributed phase-symmetrically on the phase space, and no definite phase can be attributed to it; in contrast, a coherent state has a well-defined phase.}
   \label{fig:coherent_vs_number_state}
\end{figure}

Conventionally, in the Bogoliubov approximation, the condensate is treated as being close to a state in which all the bosons occupy a coherent state of a particular condensate mode (i.e., a particular single-particle state).  When particle loss is negligible, however, the real condensate is much closer to a number state than to a coherent state (see Fig.~\hyperref[fig:coherent_vs_number_state]{\ref*{fig:coherent_vs_number_state}}).

Since a coherent state has a well-defined phase, the conventional Bogoliubov approximation breaks the $U(1)$ symmetry possessed by the condensate; consequently, a fictitious Goldstone mode~\cite{nambu_quasi-particles_1960, goldstone_field_1961} is present in the Bogoliubov Hamiltonian.  Because there is no restoring force on the Goldstone mode, the Bogoliubov ground state is not well defined; worse, the Goldstone mode causes the condensate state to deviate linearly in time from a single condensate in a coherent state (i.e., this is a secular deviation, not an oscillation).  This problem is particularly pesky when the condensate is in a trapping potential, where the Goldstone mode is a mixture of the condensate mode and modes orthogonal to it and thus cannot be removed easily.  The solution to getting rid of the unphysical Goldstone mode is to adhere to the fact that the condensate has a fixed number of particles, i.e., by using a Bogoliubov approximation where particle number is conserved.

Many authors have considered the number-conserving Bogoliubov approximation.  Girardeau and Arnowitt~\cite{girardeau_theory_1959} were the first to propose a theory for the ground state and excited states of many bosons based on a particle-number-conserving ($N$-conserving) formulation of the Bogoliubov quasiparticles.  C.~W. Gardiner~\cite{gardiner_particle-number-conserving_1997} introduced a somewhat similar approach to Girardeau and Arnowitt's, but with an emphasis on the time-dependent case; C.~W. Gardiner \emph{et al.}~\cite{gardiner_kinetics_1997} then applied this approach to the kinetics of a BEC in a trap.  Castin and Dum~\cite{castin_instability_1997, castin_low-temperature_1998, castin_bose-einstein_2001} gave a modified form of the Bogoliubov Hamiltonian where the terms that break the $U(1)$ symmetry are removed by a projection operator.  S{\o}rensen~\cite{sorensen_bogoliubov_2002} generalized the Castin-Dum result to the two-component case.  S.~A. Gardiner \emph{et al.}~\cite{gardiner_number-conserving_2007, simon_a._gardiner_number-conserving_2011, mason_number-conserving_2014} improved the Castin-Dum result in the multi-component case using an expansion in powers of the ratio of noncondensate to condensate particle numbers.  Several authors~\cite{steel_dynamical_1998, sinatra_truncated_2002} discussed the truncated Wigner approximation, which provides a way to implement a number-conserving Bogoliubov approximation in a phase-space description.

An independent approach is founded on a number-conserving BCS-like {\em ansatz\/} introduced by Leggett~\cite{leggett_bose-einstein_2001, leggett_relation_2003}.  Leggett's {\em ansatz\/} uses the state,
\begin{align}\label{eq:leggett_wave function}
\ket{\psibf_\mathrm{Legg}} \propto \Big(\,\a_0^\dagger \a_0^\dagger
+ 2\sum_{k>0} \lambda_k \a^\dagger_k \a^\dagger_{-k}\Big)^{\!N/2}\,\ket{\vac}\;,
\end{align}
as a model for analyzing the properties of the ground state of a homogeneous BEC; here $\lambda_k<1$ and $\a^\dagger_k$ creates a boson of momentum $k$.  Dziarmaga and Sacha~\cite{dziarmaga_bogoliubov_2003} generalized Leggett's {\em ansatz\/} to the inhomogeneous case while retaining a similar pair-correlated form,
\begin{align}\label{eq:pcs}
\ket{\psibf_\mathrm{pcs}}
\propto \Big(\,\a_0^\dagger \a_0^\dagger
+\sum_{m>0} \lambda_m' \a^{\dagger}_m\a^{\dagger}_m\Big)^{N/2}\,\ket{\vac}\;.
\end{align}
Here $\a^\dagger_0$ is the creation operator for the condensate mode and the creation operators $\a^{\dagger}_m$ and the real numbers $\lambda_m'$ are derived from the Bogoliobov Hamiltonian by using a singular-value decomposition.  Later, Dziarmaga and Sacha generalized their results to the time-dependent case~\cite{dziarmaga_images_2006}, where they showed that if the system starts in a Bogoliubov vacuum state, it remains in a state of the same structure.  The pair-correlated-state approach that Dziarmaga and Sacha introduced is closely related to the extended catalytic state approach discussed in the current paper.  In~\cite{jiang_bosonic_2016}, we study pair-correlated states of the form~(\ref{eq:pcs}) in great detail and demonstrate this equivalence for the case where the coefficients $\lambda_m'$ are considerably less than one---i.e., the regime where there is a single dominant condensate wavefunction; in addition, we derive analytical expressions for the physical quantities (particularly the single- and two-particle reduced density matrices) associated with pair-correlated states in the large-$N$ limit when $1-\lambda_m'\sim1/N$, a regime where more than one mode can be macroscopically occupied.

A number-conserving Bogoliubov approximation yields qualitatively different results from one that fails to conserve particle number; among these differences are the following.  Villain \emph{et al.}~\cite{villain_quantum_1997} removed the ``zero-momentum mode'' from the Boboliubov Hamiltonian and thereby showed that the collapse time of the phase of a BEC is relatively short and, in some cases, vanishes in the limit of a large number of atoms.  Danshita \emph{et al.}~\cite{danshita_collective_2005} investigated collective excitations of BECs in a box-shaped double-well trap using the number-conserving Bogoliubov approximation.  Trimborn \emph{et al.}~\cite{trimborn_exact_2008, trimborn_beyond_2009} investigated the artificial number fluctuations in methods that ignore the fixed particle number and showed that these lead to ambiguities and large deviations in the Bose-Hubbard model.  Ole\'s \emph{et al.}~\cite{oles_n-conserving_2008} predicted large density fluctuations in a two-component BEC close to the phase-separation regime using an {\em ansatz\/} in which the number of atoms in each component is fixed.  Schachenmayer \emph{et al.}~\cite{schachenmayer_atomic_2011} studied the collapse and revival of interference patterns in the momentum distribution of atoms in optical lattices using a number-projection method.  
Billam \emph{et al.}~\cite{billam_second-order_2013} went beyond the number-conserving Bogoliubov approximation and studied large depletion of the condensate by considering the coupled dynamics of the condensate and noncondensate fractions.

We return to the number-conserving Bogoliubov approximation in this paper and develop a particularly transparent method of deriving the relevant equations.  Our approach to a number-conserving Bogoliubov approximation is to ``encode'' the many-body wavefunction of the BEC in the $N$-particle sector of a state we call an {\it extended catalytic state\/} (ECS), by which we mean a coherent state for the condensate mode and a state to be determined by the dynamics for the orthogonal modes of the atoms.  Using a time-dependent interaction picture, we move the coherent state to the vacuum, thus making all the field operators formally small compared to ${N}^{1/2}$.  The resulting Hamiltonian can then be organized by powers of ${N}^{-1/2}$.  Requiring the terms of order ${N}^{1/2}$ to vanish in the interaction-picture evolution equation gives the GP equation for the condensate wavefunction.  Going to the next order in the evolution equation, $N^0$, we derive equations equivalent to those found by Castin and Dum~\cite{castin_low-temperature_1998} for a number-conserving Bogoliubov approximation.  In contrast to other approaches, ours allows one to calculate the state evolution in the Schr\"{o}dinger picture, and it also has advantages in considering higher-order corrections and extensions to multi-component cases.

In Sec.~\ref{sec:ecsbog} we introduce the ECS (Sec.~\ref{subsec:encoding}) and the interaction picture in which the condensate mode is displaced to vacuum (Sec.~\ref{subsec:ip}).  We then derive the equations that govern the ECS dynamics in this interaction picture (Sec.~\ref{subsec:bog}).  The GP equation and the Bogoliubov Hamiltonian arise naturally as we organize the interaction picture Hamiltonian by powers of ${N}^{-1/2}$.  We make the Bogoliubov Hamiltonian number conserving by adding to it an auxiliary term that does not change the dynamics in the $N$-particle sector.  Finally, we transform back to the Schr\"odinger picture and find a particularly simple form for the ECS dynamics at Bogliubov order (Sec.~\ref{subsec:bogdynamics}).  In Sec.~\ref{sec:twocomp} we generalize our approach to two-component BECs (Sec.~\ref{subsec:twocompbog}) and show how spin squeezing is generated in two-component systems (Sec.~\ref{subsec:spinsqueeze}).  Section~\ref{sec:summary} summarizes the results of the paper. This paper is based on Z. Jiang's PhD dissertation (Chapter 3) at the University of New Mexico~\cite{JiangPhD}.

\section{Extended catalytic state and number-conserving Bogoliubov approximation}
\label{sec:ecsbog}

Quantum optics teaches that coherent states are easier to deal with than number states, and what is true there is true here as well.  Indeed, the usual mean-field approximation to BEC evolution is based on the assumption that the BEC is in a coherent state of a condensate mode~\cite{pitaevskii_bose-einstein_2003}.   A problem with this approach is that the number of particles in a BEC is usually fixed, whereas coherent states are superpositions of states with different numbers of particles.  A related problem is that assigning a coherent state to a BEC breaks its phase symmetry, thus causing problems in developing the Bogoliubov approximation.

\subsection{Encoding the state of a BEC in an extended catalytic state}
\label{subsec:encoding}

Our philosophy for dealing with these problems in a BEC that has a fixed particle number $N$ is to extend the BEC state $\ket{\psibf_{\! N}}$ to a state $\ket{\psibf_\mathrm{ecs}}$, for which the condensate mode is in a coherent state, but the $N$-particle sector is the same as $\ket{\psibf_{\!N}}$ within a normalization constant.  Consider an arbitrary state $\ket{\psibf_{\!N}}$ with $N$ particles, for which we have the relative-state decomposition in the number basis of the condensate mode,
\begin{align}\label{eq:PsiN}
\ket{\psibf_N}=
\sum_{M=0}^N\, \ket{N-M}_0\otimes \ket{\varOmega_M}_\perp\;,
\qquad
\sN_\perp\, \ket{\varOmega_M}_\perp=M\, \ket{\varOmega_M}_\perp\;,
\end{align}
where the kets labeled by 0 and $\perp$ apply to the condensate mode and to all the modes orthogonal to the condensate mode, respectively.  The operator $\sN_\perp$ is the particle-number operator for the orthogonal modes.  The state $\ket{\varOmega_M}$ for the orthogonal modes, which has $M$ particles in the orthogonal modes, is not necessarily normalized.  The key to our approach is that the state~(\ref{eq:PsiN}) can be written as
\begin{align}
\ket{\psibf_N}
&=e^{\norm{\alpha}^2/2}\,\sum_{M=0}^N\, \frac{\sqrt{(N-M) !}}{\alpha^{N-M}}\; \mathcal{P}_N\Big(\ket{\alpha}_0\otimes
\ket{\varOmega_M}_\perp\Big) \nonumber \\
&=e^{\norm{\alpha}^2/2}\,\frac{\sqrt{N!}}{\alpha^N}
\;\mathcal{P}_N\Big(\ket{\alpha}_0\otimes\ket{\varOmega}_\perp\Bigr)\;,
\end{align}
where $\mathcal{P}_{N}$ is the projection operator onto the $N$-particle sector and
\begin{equation}\label{eq:Omegaperp}
\ket{\varOmega}_\perp=
\sum_{M=0}^N\;\alpha^M\sqrt{\frac{(N-M)!}{N!}}\;\ket{\varOmega_M}_\perp
\end{equation}
is an (unnormalized) state of the modes orthogonal to the condensate mode.

We now introduce the \emph{extended catalytic state},
\begin{equation}
\ket{\psibf_\mathrm{ecs}}
=\ket{\alpha}_0\otimes \ket{\varOmega}_\perp \;,\label{eq:catalytic_state}
\end{equation}
which is related to the physical state by
\begin{equation}\label{eq:psiN}
\ket{\psibf_N}=e^{\norm{\alpha}^2/2}\,\frac{\sqrt{N!}}{\alpha^N}\;
\mathcal{P}_N \ket{\psibf_\mathrm{ecs}}\;.
\end{equation}
The extended catalytic state is a direct product of a coherent state $\ket{\alpha}_0$ in the condensate mode and an unnormalized state $\ket{\varOmega}_\perp$ of the orthogonal modes.  Notice that once $\alpha$ is specified, the extended catalytic state has a one-to-one correspondence with the physical state.  The structure of the extended catalytic state allows us to study the dynamics of a BEC in the Schr\"{o}dinger picture, and we will see that the structure is preserved throughout the evolution in the Bogliubov approximation.

For a pure condensate with no depletion of the condensate mode, the modes orthogonal to the condensate mode are in vacuum, and the overall state has the form
\begin{equation}
\ket{\psibf_{N}}=\ket{N}_0\otimes \ket{\vac}_\perp\;.
\end{equation}
In this case we have
\begin{equation}
\ket{\psibf_\mathrm{ecs}}=\ket{\alpha}_0\otimes \ket{\vac}_\perp\;.
\end{equation}
Generally one expects that a dilute-gas BEC has a state close to that of a pure condensate, in which case the noncondensate state $\ket{\varOmega}_\perp$ is close to the vacuum; we want to develop an approximate description based on this expectation.  To do so, notice that the encoding into an extended catalytic state works for any value of $\alpha$.  In other words, one has the freedom to choose $\alpha$ at will; after the projection, all values of $\alpha$ yield the same physical state.   Nonetheless, we stick to the choice $|\alpha|=N^{1/2}$, for the reason that we make approximations in deriving the dynamics of $\ket{\psibf_\mathrm{ecs}}$ and the projection onto the $N$-particle sector can amplify the errors due to these approximations.  To keep these errors under control, we center the number distribution of the coherent state at the actual atomic number $N$.  The phase of $\alpha$ is yet another matter, which we discuss further below.

The BEC Hamiltonian conserves particle number and thus commutes with the particle-number operator.  As a consequence, the evolution operator $\sU(t)$ commutes with $\sP_N$, allowing us to move the evolution operator through the projection onto the $N$-particle sector so that it acts directly on the extended catalytic state:
\begin{align}\label{eq:projection}
\ket{\psibf_{\! N}(t)}
=\mathcal{U}(t)\,\ket{\psibf_{\! N}(0)}
=e^{\norm{\alpha}^2/2}\,\frac{\sqrt{N!}}{\alpha^N}\;
\mathcal{U}(t)\,\mathcal{P}_{\! N}\ket{\psibf_\mathrm{ecs}(0)}
=e^{\norm{\alpha}^2/2}\,\frac{\sqrt{N!}}{\alpha^N}\;
\mathcal{P}_{N} \ket{\psibf_\mathrm{ecs}(t)}\;.
\end{align}
To find $\ket{\psibf_{\! N}(t)}$, one solves for $\ket{\psibf_\mathrm{ecs}(t)}=\mathcal{U}(t)\,\ket{\psibf_{\!\mathrm{esc}}(0)}$ and then projects onto the $N$-particle sector.

\subsection{Interaction picture}
\label{subsec:ip}

The first step in developing the Boboliubov approximation is to go to an interaction picture in which the condensate mode is displaced from a coherent state to vacuum.  To do this, we start with a condensate mode defined by a time-dependent single-particle state $\ket{\psinot(t)}$, which has wavefunction
\begin{equation}
\psinot(\mathbf{x},t)=\braket{\mathbf{x}}{\psinot(t)}\;.
\end{equation}
The Schr\"odinger-picture field operator $\uppsi(\mathbf{x})$ satisfies the commutation relation
\begin{equation}
\big[\uppsi(\mathbf{x}),\uppsi^\dagger(\mathbf{x}')\big]=\delta(\mathbf{x}-\mathbf{x}')\;.
\end{equation}
The annihilation operator for the condensate mode is related to the Schr\"odinger-picture field operator by
\begin{equation}\label{eq:aphi}
\a_{\psinot(t)}=\int \psinot^*(\mathbf{x},t)\,\uppsi(\mathbf{x})\,\dif \mathbf{x}
=\braket{\psinot(t)}{\uppsi}=\braket{\uppsi^\dagger}{\psinot^*(t)}\;.
\end{equation}
Here, in the final two equalities, we introduce a shorthand notation for the integral as bra-ket inner products between a single-particle state and the field operator.  The creation operator for the condensate mode is
\begin{equation}\label{eq:aphidagger}
\a_{\smash{\psinot(t)}}^\dagger=\int\uppsi^\dagger(\mathbf{x})\,\psinot(\mathbf{x},t)\,\dif \mathbf{x}
=\braket{\uppsi}{\psinot(t)}=\braket{\psinot^*(t)}{\uppsi^\dagger}\;.
\end{equation}
Here and throughout this section, complex conjugation in the single-particle Hilbert space is defined relative to the position representation.  The bra-ket notation introduced here, though \emph{ad hoc}, is useful for manipulating the complicated expressions that arise as we proceed, more so once we get to the two-component case in Sec.~\ref{sec:twocomp}.  Notice that the annihilation and creation operators have two different bra-ket forms, both of which are used in our treatment.

The field operator can be written as
\begin{equation}\label{eq:fieldperp}
\uppsi(\mathbf{x})=\a_{\psinot(t)}\ssp \psinot(\mathbf{x},t)+\uppsi_{\perp}(\mathbf{x},t)\;,
\end{equation}
where $\uppsi_{\perp}(\mathbf{x},t)$ is the field operator with the condensate mode excluded.  In the Schr\"odinger picture, $\uppsi(\mathbf{x})$ is time independent, but the split between a condensate mode $\psinot(\mathbf{x},t)$ and orthogonal modes introduces time dependence because the condensate mode is changing in time; hence, both $a_{\psinot(t)}$ and $\uppsi_{\perp}(\mathbf{x},t)$ are explicitly time-dependent operators in the Schr\"odinger picture.  In terms of our shorthand notation, we can write
\begin{equation}
\ket{\uppsi_\perp(t)}
=\ket\uppsi-\ket{\psinot(t)}\a_{\psinot(t)}
=\ket\uppsi-\ket{\psinot(t)}\braket{\psinot(t)}{\uppsi}
=Q(t)\ket{\uppsi}\;,
\end{equation}
where
\begin{equation}\label{eq:Q}
Q(t)=\identity-P(t)=\identity-\proj{\psinot(t)}
\end{equation}
is the projector onto the single-particle space orthogonal to the condensate mode, with $P(t)=\proj{\psinot(t)}$ being the projector onto $\ket{\psinot(t)}$.  Notice that $Q^*(t)=\identity-P^*(t)=\identity-\proj{\psinot^*(t)}$.

The extended catalytic state for a pure condensate in the time-dependent condensate mode $\ket{\psinot(t)}$ is
\begin{equation}\label{eq:ecs_td}
\mathcal{D}\big(\alpha,\psinot(t)\big)\ket{\vac}=
\ket{\alpha,\psinot(t)}_0\otimes\ket{\vac}_\perp\;,
\end{equation}
where the displacement operator $\mathcal{D}\big( \alpha, \psinot(t) \big)$ for the condensate mode, which we usually abbreviate as $\mathcal{D}(t)$, is defined as
\begin{equation}
\mathcal{D}\big( \alpha, \psinot(t) \big)= \mathcal{D}(t)
=\exp\!\big(\alpha\ssp \a^\dagger_{\smash{\psinot(t)}} - \alpha^* \a_{\psinot(t)} \big)\;,
\end{equation}
The state~(\ref{eq:ecs_td}), which describes a pure condenstate with no depletion, is the one we perturb about in developing our approximate description.

We can now introduce the desired interaction picture as the one where the condensate mode is displaced to vacuum; i.e., states transform to
\begin{equation}\label{eq:intpic}
\ket{\psibf_\mathrm{int}(t)}
=\mathcal{D}^\dagger
\big( \alpha, \psinot(t) \big) \ket{\psibf_\mathrm{ecs}(t)}
= \mathcal{U}_\mathrm{int}(t)\,
\ket{\psibf_\mathrm{int}(0)}\;,
\end{equation}
where
\begin{equation}\label{eq:Uint}
\mathcal{U}_\mathrm{int}(t)
=\mathcal{D}^\dagger\big( \alpha, \psinot(t) \big)\,
\mathcal{U}(t)\,\mathcal{D}\big( \alpha, \psinot(0) \big)
\end{equation}
is the evolution operator in the interaction picture.  The Schr\"odinger-picture evolution operator $\mathcal{U}(t)$ obeys the Schr\"odinger equation
\begin{equation}
i\hbar\:\di{\,\mathcal{U}(t)}{t}=\mathcal{H}(t)\,\mathcal{U}(t)\;,
\end{equation}
where $\mathcal{H}(t)$ is the (possibly time-dependent) BEC Hamiltonian.  The time dependence of the condensate wavefunction $\psinot(\mathbf{x},t)$, which enters into the displacement operator $\mathcal{D}\big(\alpha, \psinot(t) \big)$ through the annihilation and creation operators, $\a_{\psinot(t)}$ and $\a_{\smash{\psinot(t)}}^\dagger$, is to be determined.

The interaction-picture evolution operator obeys the equation
\begin{align}
\begin{split}
i \hbar\: \di{\,{\mathcal U}_\mathrm{int}(t)}{t}
&=i \hbar\,\dt{\mathcal D}^\dagger(t)\, \mathcal{U}(t) \,\mathcal{D}(0)
+i\hbar\,\mathcal{D}^\dagger(t)\,\di{\,\mathcal{U}}{t}\,\mathcal{D}(0)\\[3pt]
&=\Big(\,i \hbar\,\dt{\mathcal D}^\dagger(t)\,
\mathcal{D}(t)+\mathcal{D}^\dagger(t)\,\mathcal{H}(t)\,\mathcal{D}(t)\Big)\, \mathcal{U}_\mathrm{int}(t)\;.\label{eq:evolution_Uint}
\end{split}
\end{align}
The time derivative of the displacement operator is
\begin{align}
\dt{\mathcal D}^\dagger(t)&=\di{}{t}\,\Big(e^{\alpha^* \a_{\psinot(t)}-\alpha\, \a^\dagger_{\smash{\psinot(t)}}}\Big)
=\Big(\alpha^*\,\dt{\a}_{\psinot(t)}-\alpha\,
\dt{\a}^\dagger_{\smash \psinot(t)}-\norm{\alpha}^2\,
\braket{\psinot(t)}{\dt{\psinot}(t)}\Big)\:
\mathcal{D}^\dagger(t)\;.
\end{align}
Putting this expression into Eq.~(\ref{eq:evolution_Uint}), we have
\begin{equation}
i \hbar\; \di{\,\mathcal{U}_\mathrm{int}(t)}{t}=\mathcal{H}_\mathrm{int}(t)\: \mathcal{U}_\mathrm{int}(t)\;,
\end{equation}
where the interaction-picture Hamiltonian reads
\begin{align}\label{eq:intpicH}
\mathcal{H}_\mathrm{int}(t)&=\mathord{-}i \hbar\,\Big(\,\norm{\alpha}^2\,
\braket{\psinot(t)}{\dt{\psinot}(t)}+\alpha\,
\dt{\a}^\dagger_{\smash{\psinot(t)}}-\alpha^*\, \dt{\a}_{\psinot(t)}\Big)
+\mathcal{D}^\dagger(t)\,\mathcal{H}(t)\,\mathcal{D}(t)\;.
\end{align}
Equivalently, we have
\begin{equation}
i \hbar\,\di{}{t}\ket{\psibf_\mathrm{int}(t)}=\mathcal{H}_\mathrm{int}(t)\,
\ket{\psibf_\mathrm{int}(t)}\;.
\end{equation}

In the interaction picture the field operator takes the form
\begin{equation}
\mathcal{D}^\dagger(t)\,\uppsi(\mathbf{x})\,\mathcal{D}(t)=\uppsi(\mathbf{x})+\alpha\, \psinot(\mathbf{x},t)\;.
\end{equation}
An expansion of $\mathcal{H}_\mathrm{int}(t)$ in powers of $1/|\alpha|=1/N^{1/2}$ is a good approximation as long as the field operator $\uppsi(\mathbf{x})$ is small relative to the interaction-picture displacement $\alpha\,\psinot(\mathbf{x},t)$, i.e., more formally, as long as the one-particle density matrix is small in the sense that
\begin{equation}
\rho_\mathrm{int}(\mathbf{x},\mathbf{x}')=\bra{\psibf_\mathrm{int}} \uppsi^\dagger(\mathbf{x}')\,\uppsi(\mathbf{x}) \ket{\psibf_\mathrm{int}}\sim N^0\;.
\end{equation}
This requirement is satisfied as long as the system is a condensate.  We now turn to using the expansion in powers of $1/N^{1/2}$ to derive the number-conserving Bogoliubov approximation.

\subsection{Number-conserving Bogoliubov approximation}
\label{subsec:bog}

In second-quantized form, the model Hamiltonian for the BEC is
\begin{align}\label{BECHt}
\mathcal{H}(t)
=\int\bigg[\uppsi^\dagger(\mathbf{x})\bigg(\mathord{-}\frac{{\hbar}^2}{2m}\boldsymbol{\nabla}^2 +V(\mathbf{x},t)\bigg) \uppsi(\mathbf{x})
+\frac{g}{2}\,\uppsi^\dagger(\mathbf{x})\uppsi^\dagger(\mathbf{x})\uppsi(\mathbf{x})\uppsi(\mathbf{x})\bigg]
\dif \mathbf{x} \;,
\end{align}
where the first term is the second-quantized Hamiltonian for particles trapped in a potential $V(\mathbf{x},t)$ and the second term represents the two-body scattering energy.  The only explicit time dependence in the Hamiltonian~(\ref{BECHt}) comes from a possible time dependence in the trapping potential $V(\mathbf x,t)$.  The present approach is also valid if the interaction strength $g$ is time dependent.  Such time modulation can be achieved when the $s$-wave scattering length is controlled by means of, for example, a Feshbach resonance.  For our expansion in powers of $1/\norm\alpha=1/N$ to work, we must have that $g\norm\alpha^2$ is of order $N^0$.

Going to the interaction picture, we have
\begin{align}
\mathcal{H}_\mathrm{int}(t)&= - i \hbar\,\Big(\,\norm{\alpha}^2
\braket{\psinot(t)}{\dt{\psinot}(t)}
+\alpha\,\dt{\a}^\dagger_{\smash{\psinot(t)}}-\alpha^*\,
\dt{\a}_{\psinot(t)}\,\Big)+ \mathcal{D}^\dagger(t)\,\mathcal{H}(t)\,\mathcal{D}(t)\label{eq:Hint}\\
&\simeq\norm{\alpha}^2 \int\psinot^*
\Big(\mathord{-}i \hbar\,\pa{}{t}-\frac{{\hbar}^2}{2m}\boldsymbol{\nabla}^2
+V+\frac{g}{2}\,\norm{\alpha}^2 \norm{\psinot}^2 \Big)
\psinot\: \dif \mathbf{x} \label{eq:Hint_meanfield_energy} \\
&\hspace{1.2em} +\bigg(\alpha\int
\uppsi^\dagger\Big(-i \hbar\,\pa{}{t}-\frac{{\hbar}^2}{2m}\boldsymbol{\nabla}^2 +V+g\norm{\alpha}^2\norm{\psinot}^2\Big)\psinot\,\dif \mathbf{x}+\mathrm{H.c.}\bigg) \label{eq:Hint_linear}\\
&\hspace{1.2em}
+\int\hspace{-0.25em}\bigg[\uppsi^\dagger\Big(-\frac{{\hbar}^2}{2m}\boldsymbol{\nabla}^2
+V+2 g \norm{\alpha}^2\norm{\psinot}^2\Big) \uppsi+\frac{g}{2}\Big(\alpha^2\uppsi^\dagger\uppsi^\dagger\,\psinot^2
+(\alpha^*)^2\uppsi\uppsi\,(\psinot^*)^2\Big)\bigg]\dif \mathbf{x}\;, \label{eq:Hint_quadratic}
\end{align}
where we neglect terms of order $N^{-1/2}$ or smaller. The $c$-number term~(\ref{eq:Hint_meanfield_energy}), of order $N$, is, in the time-independent case, the mean-field energy of the BEC; its only effect, in general, is to introduce a global phase, so we ignore it henceforth.

By requiring the linear term~(\ref{eq:Hint_linear}), of order ${N}^{1/2}$, to vanish, we get
\begin{equation}\label{eq:GPE}
i \hbar\,\dt{\psinot}(\mathbf{x},t)
=\bigg(\,\mathord{-}\frac{{\hbar}^2}{2m}\boldsymbol{\nabla}^2 +V(\mathbf{x},t)+g\norm{\alpha}^2
\norm{\psinot(\mathbf{x},t)}^2\,\bigg)\,\psinot(\mathbf{x},t)
=H_{\textrm{gp}}(t)\psinot(\mathbf{x},t)\;,
\end{equation}
which is the celebrated Gross-Pitaevskii equation.  The single-particle GP Hamiltonian is
\begin{equation}\label{eq:Hgp}
H_\mathrm{gp}(t)=\mathord{-}\frac{{\hbar}^2}{2m}\boldsymbol{\nabla}^2
+V(t)+g \norm{\alpha}^2\, \norm{\psinot(t)}^2\;.
\end{equation}

The structure of our approach is now clear.  By going to the interaction picture, the mean-field, Gross-Pitaevskii evolution is removed, and then by neglecting the terms of higher order than $N^0$, we are left with the quadratic Bogoliubov Hamiltonian
\begin{align}\label{eq:Hbog}
\mathcal{H}_\mathrm{bog}
=\int\bigg[\uppsi^\dagger\Big(H_{\textrm{gp}}+g\norm{\alpha}^2\norm{\psinot}^2\Big)\uppsi
+\frac{g}{2}\Big(\alpha^2\,\uppsi^\dagger \uppsi^\dagger\psinot^2
+(\alpha^*)^2\,\uppsi\uppsi (\psinot^*)^2\Big)\bigg]\dif \mathbf{x}\;,
\end{align}

To display the symplectic structure of the Bogoliubov Hamiltonian, we write it in the matrix form
\begin{equation}\label{eq:Hbog_matrixform}
\mathcal{H}_\mathrm{bog}=\half\,
\COLON
\begin{pmatrix}
\bra{\uppsi} & \bra{\uppsi^\dagger}
\end{pmatrix}
\mathsp{H}_\mathrm{bog}
\begin{pmatrix}
\ket{\uppsi}\\[5pt] \ket{\uppsi^\dagger}
\end{pmatrix}
\COLON\;,
\end{equation}
where the colons denote normal ordering of annihilation and creation operators, and the $2\times 2$ matrix $\mathsp{H}_\mathrm{bog}$ reads
\begin{equation}\label{eq:Hbog_matrix}
\mathsp{H}_\mathrm{bog}
=\begin{pmatrix}
H_\mathrm{gp} + g \norm{\alpha}^2  \norm{\psinot}^2  & g \alpha^2\,  \psinot^2 \\[6pt]
g (\alpha^*)^2  (\psinot^*)^2   & H_\mathrm{gp} + g \norm{\alpha}^2  \norm{\psinot}^2
\end{pmatrix}\;,
\end{equation}
Notice that the normal ordering has an effect only on the lower-right corner of the matrix $\mathcal{H}_\mathrm{bog}$.

As shown by Lewenstein and You~\cite{lewenstein_quantum_1996}, $\mathsp{H}_\mathrm{bog}$ has a nilpotent subspace, where phase diffusion takes place.  Such phase diffusion is not physical, but rather is a consequence of the arbitrary phase assigned to the condensate wavefunction, i.e., to $\alpha$.  This problem was addressed by introducing number-conserving approaches~\cite{girardeau_theory_1959, gardiner_particle-number-conserving_1997, castin_low-temperature_1998}.  Particularly in the work of Castin and Dum, a systematic expansion of the field operators was used in deriving the equations for the number-conserving Bogoliubov approximation.  The aim is to eliminate the artificial nilpotent subspace that gives rise to the phase diffusion.  Here we solve the same problem by introducing an additional contribution to the Hamiltonian, an auxiliary, explicitly time-dependent, Schr\"odinger-picture Hamiltonian $\sF(t)$, which does not affect the $N$-particle sector of $\ket{\psibf_\mathrm{ecs}(t)}$ and thus keeps the physical state $\ket{\psibf_{\! N}(t)}$ unchanged, i.e.,
\begin{equation}\label{eq:F_condition}
\mathcal{P}_{\! N}\, \sF(t)\,\ket{\psibf_{\mathrm{ecs}}(t)}=0\;.
\end{equation}
With this term $\sF(t)$, we can solve the phase diffusion problem by eliminating the nilpotent subspace of $\mathsp{H}_\mathrm{bog}$.

To determine the form of $\sF(t)$, we must go to the Bogoliubov level of approximation, but for now let us suppose $\sF(t)$ takes the form
\begin{align}\label{eq:F}
\sF(t)=-\frac{\eta(t)}{2}\, (\mathcal{N}-N)^2
+  \big(\alpha\ssp \a^\dagger_{\smash{\psinot(t)}}+\mathcal{N}_{\perp}(t)-N\big)\, \mathcal{F}_\perp(t)
+\big(\alpha^* \a_{\psinot(t)}-N\big)\, \mathcal{F}_\perp^\dagger(t)\;.
\end{align}
Here
\begin{equation}
\mathcal{N}=
\int \uppsi^\dagger(\mathbf{x})\uppsi(\mathbf{x})\,\dif \mathbf{x}
=\a_{\smash \psinot(t)}^\dagger\a_{\psinot(t)}
+\int \uppsi^\dagger_\perp(\mathbf{x},t)\uppsi_\perp(\mathbf{x},t)\,\dif\mathbf{x}
\end{equation}
is the total particle-number operator, and $\mathcal{N}_{\perp}=\mathcal{N}-\a_{\smash \psinot}^\dagger\a_{\psinot}$ is the particle-number operator for all the modes orthogonal to the condensate mode, i.e., the depletion number operator.  The time-dependent parameter $\eta(t)$, which is to be determined, is of order $N^{-1}$.  The operator $\mathcal{F}_\perp$, also to be determined, is of the order $N^{-1/2}$ and is a linear function of the annihilation and creation operators of the modes orthogonal to the condensate mode; it thus commutes with $\a_\psinot$ and $\a_{\smash \psinot}^\dagger$.

The first term in Eq.~(\ref{eq:F}) clearly satisfies Eq.~(\ref{eq:F_condition}). For the other two terms, we have
\begin{align}
0= \sP_N \big(\mathcal{N}-N\big)\,\ket{\alpha}_0\otimes \mathcal{F}_\perp\ket{\varOmega}_\perp
=\sP_N \big(\alpha \a_{\smash \psinot}^\dagger+\mathcal{N}_{\perp}-N\big)\mathcal{F}_\perp\,\ket{\alpha}_0\otimes \ket{\varOmega}_\perp\;,
\end{align}
and
\begin{align}
0=\big(\ssp\norm{\alpha}^2-N\big)\,\ket{\alpha}_0\otimes \mathcal{F}_\perp^\dagger\ket{\varOmega}_\perp
= \big(\alpha^* \a_{\psinot}-N\big)\mathcal{F}_\perp^\dagger\,\ket{\alpha}_0\otimes \ket{\varOmega}_\perp\;,
\end{align}
where in the first equation we use $\alpha \a_{\smash \psinot}^\dagger+\mathcal{N}_{\perp}-N=\a_{\smash \psinot}^\dagger(\alpha-a_\psinot)+\mathcal{N}-N$.
As long as the condensate mode stays in a coherent state with amplitude $\alpha$, these two terms do not affect the physical state $\ket{\psibf_{\! N}(t)}$.  We show that the condensate mode does remain in a coherent state at Bogoliubov order in Sec.~\ref{subsec:bogdynamics}.

An astute reader will have noticed that the auxiliary Hamiltonian~(\ref{eq:F}) is not Hermitian.  This is not a problem at Bogoliubov order, however, because the only nonHermitian term in $\mathcal{F}(t)$ is $\sN_\perp\sF_\perp$, which, being of order $N^{-1/2}$, can be neglected in the Bogoliubov approximation (order~$N^0$).

Going now to the interaction picture, we have
\begin{align}\label{eq:Fint}
\sF_\mathrm{int}(t)
&=\mathcal{D}^\dagger\big(\alpha, \psinot(t) \big)\,\sF(t)\,\mathcal{D}\big( \alpha, \psinot(t) \big)
=-\frac{\eta}{2}\Big(\alpha\a_{\smash \psinot}^\dagger+\alpha^*\a_\psinot+\mathcal{N}\Big)^2
+\big(\alpha\a_{\smash \psinot}^\dagger+\mathcal{N}_\perp\big)\sF_\perp
+\alpha^*\a_{\psinot}\,\mathcal{F}_\perp^\dagger\;,
\end{align}
where the identity $\norm{\alpha}^2=N$ is used to cancel several terms.  If we now discard terms of order $N^{-1/2}$ or smaller (in doing so, recall that interaction-picture field operators are order $N^0$), we obtain
\begin{align}\label{eq:FintBog}
\sF_\mathrm{int}
&=\mathord{-}\frac{\eta}{2}\,\Big(\,2\norm{\alpha}^2\,\a_{\smash \psinot}^\dagger \a_{\psinot}+\alpha^2  \a_{\smash \psinot}^\dagger \a_{\smash \psinot}^\dagger
+(\alpha^*)^2\a_\psinot\a_\psinot\Big)
+\alpha\a_{\smash \psinot}^\dagger\,\mathcal{F}_\perp+\alpha^*\a_{\psinot}\,\mathcal{F}_\perp^\dagger
-\frac{\eta}{2}|\alpha|^2\\
\begin{split}
&=\mathord{-}\frac{\eta}{2}\,\Big(
\,2\norm{\alpha}^2\,\braket{\uppsi}{\psinot}\braket{\psinot}{\uppsi}
+\alpha^2\braket{\uppsi}{\psinot}\braket{\psinot^*}{\uppsi^\dagger}
+(\alpha^*)^2\braket{\uppsi^\dagger}{\psinot^*}\braket{\psinot}{\uppsi}\Big)\\
&\qquad\qquad
+\alpha\braket{\uppsi}{\psinot}\mathcal{F}_\perp
+\alpha^*\braket{\psinot}{\uppsi}\mathcal{F}_\perp^\dagger
-\frac{\eta}{2}|\alpha|^2\;.
\end{split}
\label{eq:FintBog2}
\end{align}
Here we normally order the creation and annihilation operators of the condensate mode in preparation for incorporating $\sF_{\mathrm{int}}$ into the main Bogoliubov Hamiltonian; this normal ordering introduces the $c$-number term $-\eta|\alpha|^2/2$.  This term could be important as a second-order correction to the condensate energy, but it only adds an overall phase to the evolving quantum state, so we neglect it henceforth.  In Eq.~(\ref{eq:FintBog2}), we introduce the bra-ket notation of Eqs.~(\ref{eq:aphi}) and~(\ref{eq:aphidagger}).  The modified (number-conserving) Bogoliubov Hamiltonian then takes the form
\begin{align}\label{modHbog}
\mathcal{H}_\mathrm{ncb}
&=\mathcal{H}_\mathrm{bog}+\sF_{\mathrm{int}}\;.
\end{align}

To eliminate the phase diffusion, we choose
\begin{align}\label{eq:eta}
\eta(t)&= g \int \norm{\psinot(\mathbf{x},t)}^4\,\dif\mathbf{x}
=g\bigbra\psinot\norm{\psinot}^2\bigket\psinot
=g\bigbra{\psinot^*}\norm{\psinot}^2\bigket{\psinot^*}
=g\bigbra{\psinot^*}(\psinot^*)^2\bigket{\psinot}
=g\bigbra{\psinot}\psinot^2\bigket{\psinot^*}
\end{align}
and a Hermitian
\begin{align}
\mathcal{F}_\perp(t)
&=-g\alpha^*\!\int \psinot^*(\mathbf{x},t)\norm{\psinot(\mathbf{x},t)}^2
\uppsi_{\perp}(\mathbf{x},t)\,\dif \mathbf{x}
-g\alpha\!\int \psinot(\mathbf{x},t)\norm{\psinot(\mathbf{x},t)}^2
\uppsi_{\perp}^\dagger(\mathbf{x},t)\,\dif\mathbf{x}\\
&=-g\alpha^*\bigbra{\psinot}\norm\psinot^2Q\bigket{\uppsi}
-g\alpha\bigbra{\psinot}\psinot^2Q^*\bigket{\uppsi^\dagger}
=-g\alpha^*\bigbra{\uppsi^\dagger}Q^*(\psinot^*)^2\bigket{\psinot}
-g\alpha\bigbra{\uppsi}Q\norm\psinot^2\bigket{\psinot}\;.
\label{eq:Fperp}
\end{align}
It is now a tedious calculation to show that
\begin{align}
\begin{split}
\sF_\mathrm{int}
&=\frac{\eta}{2}\,\Big(\,2\norm{\alpha}^2\,\a_{\smash\psinot}^\dagger\a_{\psinot}
+\alpha^2\a_{\smash \psinot}^\dagger \a_{\smash \psinot}^\dagger
+(\alpha^*)^2\a_\psinot\a_\psinot\Big)\\
&\qquad-g\bigg(\Big(|\alpha|^2\a_{\smash \psinot}^\dagger+(\alpha^*)^2\a_\psinot\Big)
\int \psinot^*(\mathbf{x},t)\norm{\psinot(\mathbf{x},t)}^2\uppsi(\mathbf{x})\,\dif \mathbf{x}+\mathrm{H.c.}\bigg)
\end{split}\\[3pt]
\begin{split}
&=-\frac{g}{2}\,\Big(\,
2\norm{\alpha}^2\,\bigbra{\uppsi}P\norm\psinot^2P\bigket{\uppsi}
+(\alpha^*)^2\,\bigbra{\uppsi^\dagger}P^*(\psinot^*)^2 P\bigket{\uppsi}
+\alpha^2\,\bigbra{\uppsi}P\psinot^2 P^*\bigket{\uppsi^\dagger}
\Big)\\[2pt]
&\qquad-g\Big(
\norm{\alpha}^2\,\bigbra{\uppsi}P\norm\psinot^2Q\bigket{\uppsi}
+\norm{\alpha}^2\,\bigbra{\uppsi}Q\norm\psinot^2 P\bigket{\uppsi}\\
&\qquad\qquad+(\alpha^*)^2\,\bigbra{\uppsi^\dagger}Q^*(\psinot^*)^2P\bigket{\uppsi}
+\alpha^2\,\bigbra{\uppsi}P\psinot^2Q^*\bigket{\uppsi^\dagger}\Big)\;.
\label{eq:Fint2}
\end{split}
\end{align}

Translating this into matrix notation, we get
\begin{equation}\label{eq:Fint3}
\sF_\mathrm{int}
=\frac{1}{2}
\,\COLON
\begin{pmatrix}
\bra{\uppsi} & \bra{\uppsi^\dagger}
\end{pmatrix}
\mathsp{F}_{\mathrm{int}}
\begin{pmatrix}
\ket{\uppsi}\\[5pt] \ket{\uppsi^\dagger}
\end{pmatrix}\COLON\;,
\end{equation}
where the $2\times2$ matrix is
\begin{align}
\mathsp{F}_{\mathrm{int}}=g
\begin{pmatrix}
|\alpha|^2\big(Q|\psinot|^2Q-|\psinot|^2\big)
&\alpha^2\big(Q\psinot^2 Q^*-\psinot^2\big)\\[6pt]
(\alpha^*)^2\big(Q^*(\psinot^*)^2 Q-(\psinot^*)^2\big)
&|\alpha|^2\big(Q^*|\psinot|^2 Q^*-|\psinot|^2\big)
\end{pmatrix}\,.
\end{align}
In Eqs.~(\ref{eq:eta}), (\ref{eq:Fperp}), and (\ref{eq:Fint2}), we use the bra-ket notation, which is the easiest way to carry out the algebraic manipulations; for this purpose, it is useful to notice that $\bigbra{\uppsi}\psinot^2 P^*\bigket{\uppsi^\dagger}=\bigbra{\uppsi}P\psinot^2\bigket{\uppsi^\dagger}$ and its conjugate, $\bigbra{\uppsi^\dagger}(\psinot^*)^2 P\bigket{\uppsi}=\bigbra{\uppsi^\dagger}P^*(\psinot^*)^2\bigket{\uppsi}$.  The bra-ket manipulations generalize straightforwardly to the two-component case considered in Sec.~\ref{sec:twocomp}.

The number-conserving Bogoliubov Hamiltonian~(\ref{modHbog}) now reads
\begin{equation}\label{Hncb_matrixform}
\mathcal{H}_\mathrm{ncb}=\half\,\COLON
\begin{pmatrix}
\bra{\uppsi} & \bra{\uppsi^\dagger}
\end{pmatrix}\!
\mathsp{H}_\mathrm{ncb}
\begin{pmatrix}
\ket{\uppsi}\\[5pt] \ket{\uppsi^\dagger}
\end{pmatrix}\COLON\;,
\end{equation}
with
\begin{align}\label{eq:Hncb_single_component}
\mathsp{H}_\mathrm{ncb}
=\mathsp{H}_{\mathrm{bog}}+\mathsp{F}_{\mathrm{int}}
=\begin{pmatrix}
H_\mathrm{gp} + g \norm{\alpha}^2 Q \norm{\psinot}^2 Q & g \alpha^2\, Q \psinot^2 Q^*\\[6pt]
g (\alpha^*)^2  Q^* (\psinot^*)^2  Q & H_\mathrm{gp} + g \norm{\alpha}^2 Q^* \norm{\psinot}^2 Q^*
\end{pmatrix}\;.
\end{align}
This number-conserving Hamiltonian is the same as that found by Castin and Dum~\cite{castin_low-temperature_1998} using a systematic expansion of the field operators.  The two approaches give the same dynamics for the Bogoliubov approximation and thus are equivalent to order $N^0$.  The difference between the two approaches is that Castin and Dum derive their results in the Heisenberg picture, whereas we use the Schr\"{o}dinger picture and a closely related interaction picture.  To summarize our approach, we move the coherent state of the condensate mode to vacuum by going to a time-dependent interaction picture.  The interaction-picture Hamiltonian, when organized by powers of ${N}^{-1/2}$, gives the GP equation at order $N^{1/2}$ and the conventional Bogoliubov Hamiltonian $\mathcal H_\mathrm{bog}$ at order $N^0$.  The conventional Bogoliubov Hamiltonian suffers from the artificial problem of phase diffusion because a zero-momentum mode arises from choosing a phase for the condensate wavefunction.  Without affecting the $N$-particle sector, we remove the phase diffusion problem and derive the number-conserving Bogoliubov Hamiltonian~(\ref{eq:Hncb_single_component}) by introducing the auxiliary term~(\ref{eq:F}).

It is useful below to divide the number-conserving Bogoliubov Hamiltonian~(\ref{Hncb_matrixform}) into its two natural parts, $\mathcal{H}_\mathrm{ncb}=\mathcal{H}_\mathrm{gp}+\mathcal{\Hrem}$.  Here
\begin{align}\label{eq:Hgpcal}
\begin{split}
\mathcal{H}_\mathrm{gp}
&=\half\,\COLON
\begin{pmatrix}
\bra{\uppsi} & \bra{\uppsi^\dagger}
\end{pmatrix}\!
\mathsp{H}_\mathrm{gp}
\begin{pmatrix}
\ket{\uppsi}\\[5pt] \ket{\uppsi^\dagger}
\end{pmatrix}\COLON\\[5pt]
&=\bra{\uppsi}H_\mathrm{gp}\ket\uppsi
=\a_{\smash \psinot}^\dagger\a_\psinot\bra{\psinot}H_{\mathrm{gp}}\ket\psinot
+\a_{\smash \psinot}^\dagger\bra\psinot H_{\mathrm{gp}}\ket{\uppsi_\perp}
+\bra{\uppsi_\perp}H_{\mathrm{gp}}\ket\psinot\a_\psinot
+\bra{\uppsi_\perp}H_\mathrm{gp}\ket{\uppsi_\perp}\;,
\end{split}
\end{align}
with
\begin{equation}
\mathsp{H}_\mathrm{gp}
=\begin{pmatrix}
H_\mathrm{gp} & 0\\[6pt]
0 & H_\mathrm{gp}
\end{pmatrix}\;,
\end{equation}
is the GP part of the number-conserving Bogoliubov Hamiltonian, and
\begin{equation}
\label{eq:Kcal}
\mathcal{\Hrem}
=\half\,\COLON
\begin{pmatrix}
\bra{\uppsi} & \bra{\uppsi^\dagger}
\end{pmatrix}\!
\mathsp{\Hrem}
\begin{pmatrix}
\ket{\uppsi}\\[5pt] \ket{\uppsi^\dagger}
\end{pmatrix}\COLON
=\half\,\COLON
\begin{pmatrix}
\bra{\uppsi_\perp} & \bra{\uppsi_\perp^\dagger}
\end{pmatrix}\!
\mathsp{\Hrem}
\begin{pmatrix}
\ket{\uppsi_\perp}\\[5pt] \ket{\uppsi_\perp^\dagger}
\end{pmatrix}\COLON\;,
\end{equation}
with
\begin{align}
\label{eq:Ksp}
\mathsp{\Hrem}
&=g
\begin{pmatrix}
\norm{\alpha}^2 Q \norm{\psinot}^2 Q & \alpha^2\, Q \psinot^2 Q^*\\[6pt]
(\alpha^*)^2  Q^* (\psinot^*)^2  Q & \norm{\alpha}^2 Q^* \norm{\psinot}^2 Q^*
\end{pmatrix}
=g
\begin{pmatrix}
Q&0\\[4pt]0&Q^*
\end{pmatrix}
\begin{pmatrix}
\norm{\alpha}^2\norm{\psinot}^2 & \alpha^2\psinot^2\\[6pt]
(\alpha^*)^2(\psinot^*)^2 & \norm{\alpha}^2\norm{\psinot}^2
\end{pmatrix}
\begin{pmatrix}
Q&0\\[4pt]0&Q^*
\end{pmatrix}\;,
\end{align}
describes the additional coupling of the orthogonal modes coming from two-body scattering.  Notice that once $\mathcal{\Hrem}$ is written in the second form of Eq.~(\ref{eq:Kcal}), we can omit the projectors $Q$ and $Q^*$ from~$\mathsp{\Hrem}$.

We find it useful to introduce an orthonormal basis of single-particle states at $t=0$, $\{\ket{\psinotnot_j(0)}\}$.  We choose $\ket{\psinotnot_0(0)}=\ket{\psinot(0)}$; the $j\ge1$ states are then a complete set of modes orthogonal to $\ket{\psinot(0)}$.  We evolve all these states forward in time using the GP Hamiltonian $H_\mathrm{gp}(t)$, i.e.,
\begin{equation}\label{eq:phijt}
\ket{\psinotnot_j(t)}=U_\mathrm{gp}(t)\ket{\psinotnot_j(0)}\;,
\end{equation}
where the single-particle GP evolution operator $U_\mathrm{gp}(t)$ is the solution of
\begin{equation}\label{eq:Ugp}
i\hbar\di{U_\mathrm{gp}(t)}{t}=H_\mathrm{gp}(t)U_\mathrm{gp}(t)\;.
\end{equation}
The condensate mode $\ket{\psinot(t)}=\ket{\psinotnot_0(t)}$ satisfies Eq.~(\ref{eq:phijt}) by virtue of the GP equation~(\ref{eq:GPE}).  Notice that $U_\mathrm{gp}(t)Q(0)U_\mathrm{gp}^\dagger(t)=Q(t)$.

The corresponding annihilation and creation operators are
\begin{align}\label{eq:ajt}
\a_j(t)&=\braket{\psinotnot_j(t)}{\uppsi}=\braket{\uppsi^\dagger}{\psinotnot^*_j(t)}\;,\\
\a_j^\dagger(t)&=\braket{\psinotnot^*_j(t)}{\uppsi^\dagger}=\braket{\uppsi}{\psinotnot_j(t)}\;.
\label{eq:ajtdagger}
\end{align}
The field operators with the condensate mode excluded can be written in our bra-ket notation as
\begin{align}
\ket{\uppsi_\perp(t)}&=\sum_{j\ge1}\a_j(t)\ket{\psinotnot_j(t)}\;,\qquad
\bra{\uppsi_\perp^\dagger(t)}=\sum_{j\ge1}\a_j(t)\bra{\psinotnot^*_j(t)}\;,\\
\ket{\uppsi_\perp^\dagger(t)}&=\sum_{j\ge1}\a_j^\dagger(t)\ket{\psinotnot_j^*(t)}\;,\qquad
\bra{\uppsi_\perp(t)}=\sum_{j\ge1}\a_j^\dagger(t)\bra{\psinotnot_j(t)}\;.
\end{align}
In terms of these time-dependent single-particle states, the coupling Hamiltonian ~(\ref{eq:Kcal}), with all the time dependence indicated explicitly, takes the form
\begin{align}
\begin{split}
\mathcal{\Hrem}(t)
&=g\norm{\alpha}^2\sum_{j,k\ge1}\a_j^\dagger(t)\a_k(t)
\bigbra{\psinotnot_j(t)}\norm{\psinot}^2(t)\bigket{\psinotnot_k(t)}\\
&\quad+\frac{g}{2}\sum_{j,k\ge1}
\Big((\alpha^*)^2\a_j(t)\a_k(t)\bigbra{\psinotnot_j^*(t)}(\psinot^*)^2(t)\bigket{\psinotnot_k(t)}
+\alpha^2\a_j^\dagger(t)\a_k^\dagger(t)\bigbra{\psinotnot_j(t)}\psinot^2(t)\bigket{\psinotnot^*_k(t)}\Big)\;.
\end{split}
\label{eq:Kphimodes}
\end{align}

\subsection{Dynamics in the Bogoliubov approximation}
\label{subsec:bogdynamics}

We turn now to the dynamics of the BEC within the Bogoliubov approximation.  We begin by recalling that in the Bogoliubov approximation, an approximate interaction-picture evolution operator is constructed from the number-conserving Bogoliubov Hamiltonian~(\ref{Hncb_matrixform}), i.e.,
\begin{equation}
i \hbar\; \di{\,\mathcal{U}_\mathrm{int}(t)}{t}=\mathcal{H}_\mathrm{ncb}(t)\:\mathcal{U}_\mathrm{int}(t)\;,
\end{equation}
The corresponding interaction-picture evolution is
$\ket{\psibf_\mathrm{int}(t)}=\mathcal{U}_\mathrm{int}(t)\,\ket{\psibf_\mathrm{int}(0)}$.

Our first task is to confirm that if the condensate mode begins in a coherent state, it remains in a coherent state with the same complex amplitude $\alpha$ under the evolution of the number-conserving Bogoliubov Hamiltonian~(\ref{Hncb_matrixform}).  For this purpose, it is instructive to use Eqs.~(\ref{eq:Hgpcal}) and~(\ref{eq:Kcal}) to divide the field operators in the number-conserving Bogoliubov Hamiltonian into a contribution from the condensate mode and a contribution from the orthogonal modes:
\begin{align}
\begin{split}
\mathcal{H}_\mathrm{ncb}(t)
&=\a_{\smash \psinot}^\dagger\a_\psinot\bra{\psinot}H_{\mathrm{gp}}\ket\psinot
+\a_{\smash \psinot}^\dagger\bra\psinot H_{\mathrm{gp}}\ket{\uppsi_\perp}
+\bra{\uppsi_\perp}H_{\mathrm{gp}}\ket\psinot\a_\psinot
+\mathcal{H}_{\mathrm{ncb}\ssp\perp}\\
&=\a_{\smash \psinot}^\dagger\bra\psinot H_{\mathrm{gp}}\ket{\uppsi}
+\bra{\uppsi_\perp}H_{\mathrm{gp}}\ket\psinot\a_\psinot
+\mathcal{H}_{\mathrm{ncb}\ssp\perp}\;.
\end{split}
\label{eq:Hbogcond}
\end{align}
Here
\begin{equation}
\mathcal{H}_{\mathrm{ncb}\ssp\perp}(t)
=\half\,\COLON
\begin{pmatrix}
\bra{\uppsi_\perp} & \bra{\uppsi_\perp^\dagger}
\end{pmatrix}\!
\mathsp{H}_{\mathrm{bog}}
\begin{pmatrix}
\ket{\uppsi_\perp}\\[5pt] \ket{\uppsi_\perp^\dagger}
\end{pmatrix}\COLON
=\bra{\uppsi_\perp}H_\mathrm{gp}\ket{\uppsi_\perp}+\mathcal{\Hrem}
\end{equation}
is the Hamiltonian for the orthogonal modes; we can use $\mathsp{H}_{\mathrm{bog}}$ instead of $\mathsp{H}_\mathrm{ncb}$ because the projectors $Q$ and $Q^*$ have no effect.  Using the GP equation~(\ref{eq:GPE}), we can rewrite Eq.~(\ref{eq:Hbogcond}) as
\begin{align}
\mathcal{H}_\mathrm{ncb}
=-i\hbar\ssp \a_{\smash \psinot}^\dagger\dt\a_\psinot
+i\hbar\braket{\uppsi_\perp}{\dt\psinot}\a_\psinot
+\mathcal{H}_{\mathrm{ncb}\ssp\perp}\;,
\end{align}
from which we can immediately verify the commutator identity
\begin{align}\label{eq:comm}
\commutb{\mathcal{H}_\mathrm{ncb}(t)}{\a_{\psinot(t)}}
=i \hbar\, \dt{\a}_{\psinot(t)}\;.
\end{align}

In the Heisenberg picture, we have
\begin{equation}
i\hbar\frac{d}{dt}\,\mathcal{U}_\mathrm{int}^\dagger(t)\a_{\psinot(t)}\mathcal{U}_\mathrm{int}(t)
=\mathcal{U}_\mathrm{int}^\dagger(t)
\Big(\commutb{a_{\psinot(t)}}{\mathcal{H}_\mathrm{ncb}(t)}+i\hbar\dt\a_{\psinot(t)}\Big)
\mathcal{U}_\mathrm{int}(t)=0\;,
\end{equation}
from which we conclude that
\begin{equation}
\mathcal{U}_\mathrm{int}^\dagger(t)\a_{\psinot(t)}\mathcal{U}_\mathrm{int}(t)=\a_{\psinot(0)}\;.
\end{equation}
The conservation of $\a_\psinot$ in the Heisenberg picture implies that in the interaction picture, if the condensate mode begins in vacuum, it remains in vacuum.  Equivalent to this statement is the statement that in the Schr\"{o}dinger picture, the condensate mode is always in the coherent state $\mathcal{D}\big(\alpha,\psinot(t)\big)\ket{\mathrm{vac}}_0=
\ket{\alpha,\psinot(t)}_0$.  As a result, Eq.~(\ref{eq:F_condition}) is always satisfied, and the auxiliary Hamiltonian $\sF$ does not affect the physical state.  Notice that this means that in the extended catalytic state, the condensate mode does not become entangled with the other modes.  When we project the extended catalytic state to the $N$-particle sector to obtain the physical state of the BEC, however, entanglement makes its appearance.

To get into the Schr\"odinger picture, as we promised to do, requires some additional formal apparatus.  The effort is worthwhile, however, because it reveals the role of the GP part of the evolution at Bogoliubov order and identifies the most instructive formulation of the Bogoliobov evolution.  We begin by defining an evolution operator $\sR(t)$, which changes according to the GP part of the number-conserving Bogoliubov Hamiltonian,
\begin{equation}\label{eq:Revolution}
i \hbar\; \di{\,\sR(t)}{t}
=\mathcal{H}_\mathrm{gp}(t)\:\sR(t)
=\bra{\uppsi}H_\mathrm{gp}(t)\ket\uppsi\:\sR(t)\;.
\end{equation}
It can be seen that
\begin{align}\label{eq:PsiRevolution}
i\hbar\di{}{t}\sR^\dagger(t)\uppsi(\mathbf{x})\sR(t)
=\sR^\dagger(t)\commutb{\uppsi(\mathbf{x})}{\mathcal{H}_\mathrm{gp}(t)}\sR(t)
=\int \dif \mathbf{x'}\, H_\mathrm{gp}(\mathbf{x},\mathbf{x'},t)\,\sR^\dagger(t)\uppsi(\mathbf{x'})\sR(t)\;,
\end{align}
whose solution can be written in terms of the single-particle GP evolution operator $U_\mathrm{gp}(t)$ [see Eq.~(\ref{eq:Ugp})],
\begin{equation}
\sR^\dagger(t)\uppsi(\mathbf{x})\sR(t)
= \int \dif \mathbf{x'}\, U_\mathrm{gp}(\mathbf{x},\mathbf{x'},t)\,\uppsi(\mathbf{x'})\;;
\end{equation}
in more symbolic form, we have
\begin{equation}\label{eq:PsiRsolution}
\sR^\dagger(t)\ket{\uppsi}\ssp\sR(t)
=U_\mathrm{gp}(t)\ket{\uppsi}\;.
\end{equation}
This is the unsurprising conclusion that the field operator evolves under the GP part of Bogoliubov Hamiltonian according to the single-particle GP evolution.  Notice that $\sR(t)$ does not mix creation and annihilation operators.

Consequences of Eq.~(\ref{eq:PsiRsolution}) are the following:
\begin{align}\label{eq:aphiR}
\sR^\dagger(t)\a_{\psinot(t)}\sR(t)
&=\sR^\dagger(t)\braket{\psinot(t)}{\uppsi}\sR(t)
=\bra{\psinot(t)}U_\mathrm{gp}(t)\ket{\uppsi}
=\braket{\psinot(0)}{\uppsi}=\a_{\psinot(0)}\;,\\
\sR^\dagger(t)\ket{\uppsi_\perp(t)}\sR(t)
&=Q(t)\sR^\dagger(t)\ket{\uppsi}\sR(t)
=Q(t)U_\mathrm{gp}(t)\ket{\uppsi}
=U_\mathrm{gp}(t)Q(0)\ket{\uppsi}
=U_\mathrm{gp}(t)\ket{\uppsi_\perp(0)}\;.
\label{eq:PsiperpR}
\end{align}
Equation~(\ref{eq:aphiR}) says that $\a_\psinot$ is conserved under the GP evolution.  More generally, Eq.~(\ref{eq:PsiperpR}) is the statement that the annihilation operators for all the orthogonal modes propagated using the single-particle GP Hamiltonian [see~(\ref{eq:ajt})] are also conserved, just as in Eq.~(\ref{eq:aphiR}), i.e.,
\begin{equation}
\sR^\dagger(t)\a_j(t)\sR(t)=\a_j(0)\equiv\a_j\;.
\label{eq:ajevolution}
\end{equation}

The evolution operator $\sR(t)$ is very close to being a formal device: it translates between the natural modal descriptions that apply at different times as the single-particle states evolve under the single-particle GP Hamiltonian.  Indeed, it is easy to see that
\begin{align}\label{eq:RFock}
\sR(t)\bigket{n_0,\psinot(0);n_1,\psinotnot_1(0);n_2,\psinotnot_2(0);\ldots}
=\bigket{n_0,\psinot(t);n_1,\psinotnot_1(t);n_2,\psinotnot_2(t);\ldots}\;,
\end{align}
where $\ket{n_j,\psinotnot_j(t)}=[\a_j^\dagger(t)]^{n_j}\ket{\vac}/\sqrt{\vphantom{i}n_j}$ is the state with $n_j$ particles in the single-particle state $\ket{\psinotnot_j(t)}$.  This gives us $\sR(t)$ as an explicit basis transformation:
\begin{align}\label{eq:Rexplicit}
\sR(t)
=\sum_{n_0,n_1,n_2,\ldots}
\bigket{n_0,\psinot(t);n_1,\psinotnot_1(t);n_2,\psinotnot_2(t);\ldots}
\bigbra{n_0,\psinot(0);n_1,\psinotnot_1(0);n_2,\psinotnot_2(0);\ldots}\;.
\end{align}
We find it useful to have available the restriction of $\sR(t)$ to the orthogonal modes:
\begin{align}
\sRperp(t)
=\sum_{n_1,n_2,\ldots}
\bigket{n_1,\psinotnot_1(t);n_2,\psinotnot_2(t);\ldots}\bigbra{n_1,\psinotnot_1(0);n_2,\psinotnot_2(0);\ldots}\;.
\end{align}

All this suggests going to an interaction picture relative to the GP part of the Hamiltonian and solving for the evolution operator
\begin{equation}
\mathcal{V}(t)=\sR^\dagger(t)\,\mathcal{U}_\mathrm{int}(t)\;,
\end{equation}
which obeys the evolution equation
\begin{equation}\label{eq:Vevolution}
i\hbar\di{\,\mathcal{V}(t)}{t}=
\tilde{\mathcal{\Hrem}}(t)
\mathcal{V}(t)\;,
\end{equation}
where
\begin{align}\label{eq:RGR}
\tilde{\mathcal{\Hrem}}(t)
=\sR^\dagger(t)\mathcal{\Hrem}(t)\sR(t)
=\half\,\COLON
\begin{pmatrix}
\bra{\uppsi_\perp(0)} & \bra{\uppsi_\perp^\dagger(0)}
\end{pmatrix}\!
\tilde{\mathsp{\Hrem}}(t)
\begin{pmatrix}
\ket{\uppsi_\perp(0)}\\[5pt] \ket{\uppsi_\perp^\dagger(0)}
\end{pmatrix}\COLON\;.
\end{align}
Here the matrix of symplectic structure is
\begin{align}
\begin{split}
\tilde{\mathsp{\Hrem}}(t)
&=
\begin{pmatrix}
U_\mathrm{gp}^\dagger(t)&0\\[2pt]0&U_\mathrm{gp}^T(t)
\end{pmatrix}
\mathsp{\Hrem}(t)
\begin{pmatrix}
U_\mathrm{gp}(t)&0\\[2pt]0&U^*_\mathrm{gp}(t)
\end{pmatrix}\\[4pt]
&=g
\begin{pmatrix}
Q(0)&0\\[4pt]0&Q^*(0)
\end{pmatrix}
\begin{pmatrix}
U_\mathrm{gp}^\dagger(t)&0\\[2pt]0&U_\mathrm{gp}^T(t)
\end{pmatrix}\\[2pt]
&\qquad\times
\begin{pmatrix}
\norm{\alpha}^2\norm{\psinot}^2(t) & \alpha^2\psinot^2(t)\\[6pt]
(\alpha^*)^2(\psinot^*)^2(t) & \norm{\alpha}^2\norm{\psinot}^2(t)
\end{pmatrix}
\begin{pmatrix}
U_\mathrm{gp}(t)&0\\[2pt]0&U^*_\mathrm{gp}(t)
\end{pmatrix}
\begin{pmatrix}
Q(0)&0\\[4pt]0&Q^*(0)
\end{pmatrix}
\;.
\end{split}
\label{eq:tildeK}
\end{align}
The projectors $Q(0)$ and $Q^*(0)$ can be omitted when the matrix $\tilde{\mathsp{\Hrem}}(t)$ is inserted into Eq.~(\ref{eq:RGR}).

We can get a better idea of what the Hamiltonian~(\ref{eq:RGR}) means by writing it in terms of the time-dependent single-particle states considered in Eqs.~(\ref{eq:phijt}), (\ref{eq:ajt}), and~(\ref{eq:ajtdagger}):
\begin{align}\label{eq:tildeKtwo}
\begin{split}
\tilde{\mathcal{\Hrem}}(t)
&=g\norm{\alpha}^2\sum_{j,k\ge1}
\a_j^\dagger\a_k
\bigbra{\psinotnot_j(0)}U_\mathrm{gp}^\dagger(t)\norm{\psinot}^2(t)U_\mathrm{gp}(t)\bigket{\psinotnot_k(0)}\\
&\quad+\frac{g}{2}\sum_{j,k\ge1}\Big(
(\alpha^*)^2\a_j\a_k\bigbra{\psinotnot_j^*(0)}U_\mathrm{gp}^T(t)(\psinot^*)^2(t)
U_\mathrm{gp}(t)\bigket{\psinotnot_k(0)}
\\&\quad\hspace{6em}
+\alpha^2\a_j^\dagger\a_k^\dagger\bigbra{\psinotnot_j(0)}U^\dagger_\mathrm{gp}(t)\psinot^2(t)
U^*_\mathrm{gp}(t)\bigket{\psinotnot^*_k(0)}\Big)\;.
\end{split}\\[6pt]
\begin{split}
&=g\norm{\alpha}^2\sum_{j,k\ge1}
\a_j^\dagger\a_k
\bigbra{\psinotnot_j(t)}\norm{\psinot}^2(t)\bigket{\psinotnot_k(t)}\\
&\quad+\frac{g}{2}\sum_{j,k\ge1}\Big(
(\alpha^*)^2\a_j\a_k\bigbra{\psinotnot_j^*(t)}(\psinot^*)^2(t)\bigket{\psinotnot_k(t)}
+\alpha^2\a_j^\dagger\a_k^\dagger\bigbra{\psinotnot_j(t)}\psinot^2(t)\bigket{\psinotnot^*_k(t)}\Big)\;.
\end{split}\label{eq:tildeKthree}
\end{align}
The form~(\ref{eq:tildeKthree}) can be obtained directly from applying Eq.~(\ref{eq:ajevolution}) to Eq.~(\ref{eq:Kphimodes}), or it can be obtained by the route through Eq.~(\ref{eq:tildeKtwo}), which shows that the role of the single-particle GP evolution operators is to transform the coupling matrix elements into the time-dependent basis $\{\ket{\psinotnot_j(t)}\}$.

We can now write the Schr\"odinger-picture evolution operator as
\begin{equation}
\mathcal{U}(t)
=\mathcal{D}\big(\alpha,\psinot(t)\big)
\mathcal{U}_\textrm{int}(t)
\mathcal{D}^\dagger\big(\alpha,\psinot(0)\big)
=\mathcal{D}\big(\alpha,\psinot(t)\big)
\sR(t)\mathcal{V}(t)
\mathcal{D}^\dagger\big(\alpha,\psinot(0)\big)\;.
\end{equation}
Using Eq.~(\ref{eq:aphiR}), we have $\sR^\dagger(t)\mathcal{D}\big(\alpha,\psinot(t)\big)\sR(t)=\mathcal{D}\big(\alpha,\psinot(0)\big)$, and noting that $\mathcal{V}(t)$ only acts on the orthogonal modes, we can remove the displacement operators from the evolution operator, obtaining
\begin{equation}\label{eq:URV}
\mathcal{U}(t)
=\sR(t)\mathcal{V}(t)\;.
\end{equation}
The upshot of all this is that the Schr\"odinger-picture evolution involves, first, evolution of the orthogonal modes, with fixed creation and annihilation operators, under the Hamiltonian~(\ref{eq:tildeKthree}) and, second, translation of the mode structure to the time-dependent modes evolved using the single-particle GP Hamiltonian.  The displacement of the condensate mode to vacuum is, as we anticipated, a formal device for developing the expansion in powers of $1/\norm{\alpha}=1/N^{1/2}$; it disappears from the final Schr\"odinger-picture evolution.

Suppose now that, in accordance with our general assumptions, the initial extended catalytic state is $\ket{\psibf_\mathrm{ecs}(0)}=\ket{\alpha,\psinot(0)}_0\otimes\ket{\varOmega(0)}_\perp$.  Then $\mathcal{V}(t)\ket{\psibf_\mathrm{ecs}(0)}=\ket{\alpha,\psinot(0)}_0\otimes\mathcal{V}(t)\ket{\varOmega(0)}_\perp$, since $\mathcal{V}(t)$ only acts on the orthogonal modes.  The operator $\sR(t)$ translates this state to the modes that apply at time $t$, giving
\begin{equation}
\ket{\psibf_\mathrm{ecs}(t)}=\ket{\alpha,\psinot(t)}_0\otimes\ket{\varOmega(t)}_\perp\;,
\end{equation}
where
\begin{equation}
\ket{\varOmega(t)}_\perp=\sRperp(t)\mathcal{V}(t)\ket{\varOmega(0)}_\perp\;.
\end{equation}

To find the physical state at time~$t$, one projects the extended catalytic state onto the $N$-particle sector, as specified by Eq.~(\ref{eq:projection}), which gives
\begin{equation}
\ket{\psibf_N(t)}=\sum_{M=0}^N\bigket{N-M,\psinot(t)}_0\otimes\ket{\varOmega_M(t)}_\perp\;,
\end{equation}
where
\begin{align}
\begin{split}
\ket{\varOmega_M(t)}_\perp
&=\frac{1}{\alpha^M}\sqrt{\frac{N!}{(N-M)!}}\,\mathcal{P}_{\perp,M}(t)\ket{\varOmega(t)}_\perp\\
&=\sum_{M'=0}^N\alpha^{M'-M}\sqrt{\frac{(N-M')!}{(N-M)!}}\,\mathcal{P}_{\perp,M}(t)
\sRperp(t)\mathcal{V}(t)\ket{\varOmega_{M'}(0)}_\perp\;.
\end{split}
\end{align}
Here $\mathcal{P}_{\perp,M}(t)$ projects onto the $M$-particle sector of the modes orthogonal to the condensate mode at time~$t$.  Notice that $M'$ does not have to equal $M$ because $\mathcal{V}(t)$ is not number conserving.

\section{Two-component BECs}
\label{sec:twocomp}

In Sec.~\ref{sec:ecsbog} we discussed how to derive the number-conserving Bogoliubov approximation for a single-component BEC by going to an interaction picture where the condensate mode is displaced to vacuum.  In this section we show that it is a simple task to generalize our method to multi-component BECs.  We do the two-component case as an example, but the generalization to many components is straightforward.

We are certainly not the first to consider a number-conserving Bogoliubov approximation for the multi-component case.  S{\o}rensen~\cite{sorensen_bogoliubov_2002} generalized the Castin-Dum result to the two-component case, and this facilitated discussions on spin squeezing in BECs~\cite{esteve_squeezing_2008, grond_optimizing_2009, riedel_atom-chip-based_2010, gross_nonlinear_2010}.  S.~A. Gardiner \emph{et al.}~\cite{gardiner_number-conserving_2007, simon_a._gardiner_number-conserving_2011, mason_number-conserving_2014} improved the Castin-Dum and S{\o}rensen results by using an expansion in powers of the ratio of noncondensate to condensate particle numbers, which is advantageous for large depletion.  Compared to these previous studies, our approach for the multi-component case is distinguished mainly by the ability to carry over the single-component case with very little modification, essentially a generalization to a spinor notation for the several components.

\subsection{Number-conserving Bogoliubov approximation for two-component BECs}
\label{subsec:twocompbog}

In the two-component case the condensate wavefunction, which is generally a single-particle state that is entangled between the translational and internal degrees of freedom, takes the form
\begin{align}
\ket{\psinot(t)}
=\frac{1}{\alpha}\sum_\sigma\alpha_\sigma(t)\ket{\psinot_\sigma(t)}\otimes\ket\sigma
=\frac{1}{\alpha}\,\Big(\,\alpha_1(t)\,
\ket{\psinot_1(t)}\otimes \ket{1}+\alpha_2(t)\, \ket{\psinot_2(t)}\otimes \ket{2}\,\Big)\;,
\label{eq:condensate_state_two_component}
\end{align}
where $\sigma$, which takes on values 1 and 2 in the two-component case, labels the hyperfine levels and where $\norm{\alpha_1}^2+\norm{\alpha_2}^2=\norm{\alpha}^2=N$, with $N$ being the total number of particles. The states $\ket1$ and $\ket2$ are internal states of the bosonic atoms, which we refer to as hyperfine levels because that would be a typical situation in a dilute-gase {BEC}.  In the subspace spanned by $\ket{\psinot_1}\otimes \ket{1}=\ket{1,\psinot_1}$ and $\ket{\psinot_2}\otimes \ket{2}=\ket{2,\psinot_2}$, the single-particle state that is orthogonal to the condensate mode is
\begin{equation}\label{eq:barpsinot}
\ket{\bar\psinot(t)}
=\frac{1}{\alpha^*}\,\Big(\,\alpha_2^*(t)\, \ket{\psinot_1(t)}\otimes \ket{1}-\alpha_1^*(t)\, \ket{\psinot_2(t)}\otimes \ket{2}\,\Big)\;.
\end{equation}
Notice that
\begin{align}
\ket{1,\psinot_1(t)}
&\equiv\ket{\psinot_1(t)}\otimes\ket1
=\bigg(\frac{\alpha_1^*(t)}{\alpha^*}\ket{\psinot(t)}+\frac{\alpha_2(t)}{\alpha}\ket{\bar\psinot(t)}\bigg)\;,\\
\ket{2,\psinot_2(t)}
&\equiv\ket{\psinot_2(t)}\otimes\ket2
=\bigg(\frac{\alpha_2^*(t)}{\alpha^*}\ket{\psinot(t)}-\frac{\alpha_1(t)}{\alpha}\ket{\bar\psinot(t)}\bigg)\;.
\end{align}

The field operator that destroys a particle in internal level~$\sigma$ at position $\mathbf x$ is $\uppsi_\sigma(\mathbf x)$.  In our shorthand bra-ket notation for field operators, we have
\begin{equation}
\uppsi_\sigma(\mathbf x)=\braket{\mathbf x}{\uppsi_\sigma}=\braket{\sigma,\mathbf x}{\uppsi}\;.
\end{equation}
In the final form, we extend our notation by introducing a total field operator
\begin{equation}
\ket{\uppsi}=\sum_\sigma\,\ket{\uppsi_\sigma}\ket\sigma\;,
\end{equation}
which is a spinor field operator, including both spatial and internal degrees of freedom.  It gives the hyperfine-level field operators according to $\braket{\sigma}{\uppsi}=\ket{\uppsi_\sigma}$; notice that since $\uppsi_\sigma^\dagger(\mathbf x)=\braket{\uppsi_\sigma}{\mathbf x}=\braket{\uppsi}{\sigma,\mathbf x}$, we also have $\braket{\uppsi}{\sigma}=\bra{\uppsi_\sigma}$.  The spinor representation is
\begin{equation}
\uppsi(\mathbf x)=\braket{\mathbf x}{\uppsi}=\sum_\sigma\uppsi_\sigma(\mathbf x)\ket\sigma\;.
\end{equation}
We can also write $\uppsi_\sigma(\mathbf x)=\braket{\uppsi_\sigma^\dagger}{\mathbf x}=\braket{\uppsi^\dagger}{\sigma,\mathbf x}$ and $\uppsi_\sigma^\dagger(\mathbf x)=\braket{\mathbf x}{\uppsi_\sigma^\dagger}=\braket{\sigma,\mathbf x}{\uppsi^\dagger}$.

The annihilation and creation operators that destroy or create a particle in internal level $\sigma$ with spatial wavefunction $\psi(\mathbf x)$ are
\begin{align}\label{eq:achi}
b_{\sigma,\psi}&=\int \psi^*\nsp(\mathbf{x})\,\uppsi_\sigma(\mathbf{x})\,\dif\mathbf{x}
=\braket{\psi}{\uppsi_\sigma}=\braket{\sigma,\psi}{\uppsi}
\,,\\[3pt]
b_{\sigma,\psi}^\dagger&=\int \psi(\mathbf{x})\,\uppsi_\sigma^\dagger(\mathbf{x})\,\dif\mathbf{x}
=\braket{\uppsi_\sigma}{\psi}=\braket{\uppsi}{\sigma,\psi}
\;.
\end{align}
The annihilation operators for the entangled states $\ket\psinot$ and $\ket{\bar\psinot}$ are thus
\begin{align}\label{eq:ab}
\a_{\psinot}&=\braket{\psinot}{\uppsi}
=\frac{1}{\alpha^*}\,\Big(\alpha_1^* \braket{\psinot_1}{\uppsi_1}+\alpha_2^*\braket{\psinot_2}{\uppsi_2}\Big)
=\frac{1}{\alpha^*}\,
\big(\alpha_1^*\ssp b_{1,\psinot_1}+\alpha_2^*\ssp b_{2,\psinot_2}\big)\;,\\[4pt]
\bar{\a}_{\psinot}&=\braket{\bar\psinot}{\uppsi}
=\frac{1}{\alpha}\,\Big(\alpha_2\braket{\psinot_1}{\uppsi_1}-\alpha_1\braket{\psinot_2}{\uppsi_2}\Big)
=\frac{1}{\alpha}\,\big(\alpha_2\ssp b_{1,\psinot_1}-\alpha_1\ssp b_{2,\psinot_2}\big)\;.
\label{eq:abarb}
\end{align}
The field operator for the atoms in hyperfine level~$\sigma$ can be written as
\begin{align}
\uppsi_\sigma(\mathbf x)
&=b_{\sigma,\psinot_\sigma(t)}\psinot_\sigma(\mathbf x,t)+\uppsi_{\sigma\perp}(\mathbf x,t)\;.
\end{align}
The total field operator can be written in a variety of forms,
\begin{align}
\ket{\uppsi}
&=\sum_\sigma b_{\sigma,\psinot_\sigma(t)}\ket{\psinot_\sigma(t)}\otimes\ket\sigma+\ket{\uppsi_\doubleperp(t)}\\
&=\a_{\psinot(t)}\ket{\psinot(t)}+\bar\a_{\psinot(t)}\ket{\bar\psinot(t)}+ \ket{\uppsi_\doubleperp(t)}\label{eq:uppsidoubleperp}\\[7pt]
&=\a_{\psinot(t)}\ket{\psinot(t)}+\ket{\uppsi_\perp(t)}\label{eq:uppsiperp}\;,
\end{align}
where
\begin{equation}
\ket{\uppsi_\doubleperp(t)}=\sum_\sigma\,\ket{\uppsi_{\sigma\perp}(t)}\ket\sigma
\end{equation}
is the total field operator with modes $\ket\psinot$ and $\ket{\bar\psinot}$ removed and
\begin{align}\label{eq:psiperptwo}
\ket{\uppsi_\perp(t)}
=\bar\a_{\psinot(t)}\ket{\bar\psinot(t)}+\ket{\uppsi_\doubleperp(t)}
=\ket\uppsi-\a_{\psinot(t)}\ket{\psinot(t)}
=Q(t)\ket{\uppsi}
\end{align}
is the total field operator with only the condensate mode removed.  The projectors onto and orthogonal to the condensate mode,
\begin{align}
P(t)=\proj{\psinot(t)}\;,\qquad
Q(t)=\identity-P(t)\;,
\end{align}
are defined as in the single-component case [cf.~Eq.~(\ref{eq:Q})].  By using our bra-ket shorthand, all the manipulations for two components can be made identical to that for a single component.

Just as in the single-component case, we perturb about the extended catalytic state for a pure condensate that is in a coherent state for the condensate mode:
\begin{equation}
\mathcal{D}\big(\alpha,\psinot(t)\big)\ket{\vac}=
\ket{\alpha,\psinot(t)}_0\otimes\ket{\vac}_\perp\;,
\end{equation}
The physical state is obtained by projecting onto the $N$-particle sector.

In the two-component case the model Hamiltonian for the $N$ atoms is
\begin{align}
\mathcal{H}(t)
=\sum_\sigma\int \uppsi^\dagger_\sigma\Big(\mathord{-}
\frac{{\hbar}^2}{2m_\sigma}\boldsymbol{\nabla}^2 +V_\sigma(t)\Big)
\uppsi_\sigma\,\dif \mathbf{x}
+\sum_{\sigma,\tau}\hbar\omega_{\sigma\tau}\!
\int\uppsi^\dagger_\sigma\uppsi_\tau\,\dif \mathbf{x}
+\frac{1}{2}\,\sum_{\sigma,\tau}g_{\sigma\tau}
\int\uppsi^\dagger_\sigma\uppsi^\dagger_\tau\uppsi_\tau\uppsi_\sigma\,\dif\mathbf{x}\;.\label{eq:H_two_component}
\end{align}
The diagonal terms of the Hermitian matrix $\hbar\omega_{\sigma\tau}$ give the energies of the internal levels, and the off-diagonal terms give the single-particle coupling between the two levels.  The real, symmetric matrix $g_{\sigma\tau}$ describes the scattering of the atoms in each component off one another and the cross-scattering between components.  Since the single-particle terms are trivial to treat, the really new effect comes from the cross scattering described by $g_{12}$.

The next step is to go to the interaction picture where the condensate mode is displaced to vacuum, just as in Eq.~(\ref{eq:intpic}).  In this interaction picture, the field operators transform according~to
\begin{equation}
\mathcal{D}^\dagger\big( \alpha, \psinot(t) \big)\,\uppsi_\sigma(\mathbf{x})\,
\mathcal{D}\big( \alpha, \psinot(t) \big)
=\alpha_\sigma(t)\, \psinot_\sigma(\mathbf{x},t)+\uppsi_\sigma(\mathbf{x})\;,
\end{equation}
thus allowing an expansion in powers of $1/{N}^{1/2}=1/|\alpha|$.  We can write this transformation more abstractly as
\begin{equation}
\bigket{\mathcal{D}^\dagger\big(\alpha,\psinot(t)\big)\,\uppsi\,
\mathcal{D}\big(\alpha,\psinot(t)\big)}
=\alpha\ket{\psinot(t)}+\ket\uppsi\;,
\end{equation}
The interaction-picture Hamiltonian, as in Eq.~(\ref{eq:intpicH}), is given by
\begin{align}
\mathcal{H}_\mathrm{int}(t)&=- i \hbar\, \Big(\, \norm{\alpha}^2
\braket{\psinot(t)}{\dt{\psinot}(t)}+\alpha\,
\braket{\uppsi}{\dt{\psinot}(t)}
-\,\alpha^*\, \braket{\dt{\psinot}(t)}{\uppsi}\,\Big)+ \mathcal{D}^\dagger(t)\,\mathcal{H}(t)\,\mathcal{D}(t)\;.
\end{align}
The time derivative of the condensate state is
\begin{equation}
\ket{\dt{\psinot}(t)}
=\frac{1}{\alpha}\left(
\frac{d}{dt}\big(\alpha_1(t)\,\ket{\psinot_1(t)}\big)\otimes\ket1+
\frac{d}{dt}\big(\alpha_2(t)\,\ket{\psinot_2(t)}\big)\otimes\ket2
\right)\;.
\end{equation}
Putting all this together, we get the interaction-picture Hamiltonian to Bogoliubov order, i.e., order $N^0$,
\begin{align}
\mathcal{H}_\mathrm{int}(t)
&=\int\sum_\sigma\alpha_\sigma^*\psinot_\sigma^*
\bigg[\bigg(
\mathord{-}i\hbar \pa{}{t}+H_\sigma+\half\,\sum_\tau g_{\sigma\tau}\norm{\alpha_\tau}^2\,\norm{\psinot_\tau}^2
\bigg)\alpha_\sigma\psinot_\sigma+\sum_\tau\hbar\omega_{\sigma\tau}\alpha_\tau\psinot_\tau
\bigg]\,\dif\mathbf{x}\\
&\quad+\int\Bigg(\sum_\sigma\uppsi_\sigma^\dagger
\bigg[\bigg(
\mathord{-}i\hbar \pa{}{t}+H_\sigma+\sum_\tau g_{\sigma\tau}\norm{\alpha_\tau}^2\,\norm{\psinot_\tau}^2
\bigg)\alpha_\sigma\psinot_\sigma+\sum_\tau\hbar\omega_{\sigma\tau}\alpha_\tau\psinot_\tau
\bigg]
+\mathrm{H.c.}\Bigg)\,\dif\mathbf{x}\\
\begin{split}
&\quad+\int\bigg[\sum_\sigma\uppsi_\sigma^\dagger
\bigg(H_\sigma+\sum_\tau g_{\sigma\tau}\norm{\alpha_\tau}^2\norm{\psinot_\tau}^2\bigg)\uppsi_\sigma+\sum_{\sigma,\tau}\uppsi_\sigma^\dagger
\Big(\hbar\omega_{\sigma\tau}+g_{\sigma\tau}\alpha_\sigma\psinot_\sigma\alpha_\tau^*\psinot_\tau^*\Big)
\uppsi_\tau\\
&\hspace{5em}
+\half\sum_{\sigma,\tau}\Bigl(\uppsi_\sigma^\dagger\uppsi_\tau^\dagger
g_{\sigma\tau}\alpha_\sigma\psinot_\sigma\alpha_\tau\psinot_\tau
+\mathrm{H.c.}\Big)\bigg]
\dif \mathbf{x}\;,
\end{split}
\label{eq:Hbog_two_component}
\end{align}
where the single-body translational Hamiltonians are
\begin{equation}
H_\sigma=\mathord{-}\frac{{\hbar}^2}{2m_\sigma}\boldsymbol{\nabla}^2 +V_\sigma\;.
\end{equation}

Just as for a single component, we can neglect the $c$-number, mean-field-energy term.  By requiring the term of order $N^{1/2}=\norm{\alpha}$ to vanish, we get a pair of coupled GP equations,
\begin{equation}
\mbox{{\small $\displaystyle{\bigg(
\mathord{-}i\hbar\pa{}{t}+H_\sigma+\sum_\tau g_{\sigma\tau}\norm{\alpha_\tau}^2\,\norm{\psinot_\tau}^2
\bigg)
\alpha_\sigma\psinot_\sigma
+\sum_\tau\hbar\omega_{\sigma\tau}\alpha_\tau\psinot_\tau
=0\;.}$}}
\end{equation}
Notice that these are best thought of as coupled equations for the unnormalized wavefunctions, $\alpha_1\psinot_1$ and $\alpha_2\psinot_2$.  It is often convenient to have the two GP equations written out separately as
\begin{align}\label{eq:GP_two_component}
\begin{split}
\Big(\mathord{-}i \hbar \pa{}{t}+H_{\mathrm{gp}}^{(1)}\,\Big)\,\alpha_1 \psinot_1+\hbar\omega_{12}\,\alpha_2\psinot_2=0\;,\\[6pt]
\Big(\mathord{-}i \hbar \pa{}{t}+H_{\mathrm{gp}}^{(2)}\,\Big)\,\alpha_2 \psinot_2+\hbar\omega_{21}\,\alpha_1\psinot_1=0\;,
\end{split}
\end{align}
where the GP Hamiltonians are
\begin{gather}\label{eq:GP1}
H_\mathrm{gp}^{(1)}
=H_1+\hbar\omega_{11}+g_{1\nsp 1}\norm{\alpha_1}^2 \norm{\psinot_1}^2
+g_{1\nsp 2}\norm{\alpha_2}^2\norm{\psinot_2}^2\;,\\[6pt]
H_\mathrm{gp}^{(2)}
=H_2+\hbar\omega_{22}+g_{22}\norm{\alpha_2}^2\norm{\psinot_2}^2
+g_{2\nsp 1}\norm{\alpha_1}^2\norm{\psinot_1}^2
\label{eq:GP2}
\end{gather}
(remember that $\omega_{21}=\omega_{12}^*$ and $g_{21}=g_{12}$).  It is also convenient
to make the equations compact by writing them in terms of spinors relative to the two hyperfine levels so that we can take advantage of our bra-ket notation,
\begin{equation}\label{eq:spinorGP}
\Big(\mathord{-}i\hbar\pa{}{t}+H_{\mathrm{gp}}\,\Big)
\begin{pmatrix}
\alpha_1\psinot_1\\\alpha_2\psinot_2
\end{pmatrix}
=0\;,
\end{equation}
where
\begin{align}
H_\mathrm{gp}&=\begin{pmatrix}
H_\mathrm{gp}^{(1)} &
\hbar\omega_{12}\\[3pt]
\hbar\omega_{21} &
H_\mathrm{gp}^{(2)}
\end{pmatrix}
=H_\mathrm{gp}^{(1)}\proj{1}+H_\mathrm{gp}^{(2)}\proj{2}+
\hbar\omega_{12}\ket1\bra2+\hbar\omega_{21}\ket2\bra1\;.\label{eq:Hgpmatrix}
\end{align}
Recognizing that the spinor in Eq.~(\ref{eq:spinorGP}) is the spinor representation of the state $\alpha\ket\psinot$,
we can write the coupled GP equations in the very compact form
\begin{equation}\label{eq:GPcompact}
\Big(\mathord{-}i\hbar\pa{}{t}+H_{\mathrm{gp}}\,\Big)\ket\psinot=0\;,
\end{equation}
where it is assumed, as our formalism requires, that $\alpha$ does not change in time.

The coupled GP equations~(\ref{eq:GP1}) and~(\ref{eq:GP2}) imply that
\begin{align}\label{eq:normchange}
\di{}{t}\big(|\alpha_1|^2\vecnorm{\psinot_1}\big)
&=-\di{}{t}\big(|\alpha_2|^2\vecnorm{\psinot_2}\big)
=2\,{\rm Im}\big(\omega_{12}\alpha_1^*\alpha_2\braket{\psinot_1}{\psinot_2}\big)\;,\\
i\hbar\di{}{t}\big(\alpha_1^*\alpha_2\braket{\psinot_1}{\psinot_2}\big)
&=\alpha_1^*\alpha_2\bigbra{\psinot_1}\big(H_{\mathrm{gp}}^{(2)}-H_{\mathrm{gp}}^{(1)}\big)\bigket{\psinot_2}
+\hbar\omega_{12}^*\big(\norm{\alpha_1}^2\vecnorm{\psinot_1}-\norm{\alpha_2}^2\vecnorm{\psinot_2}\big)\;.
\label{eq:overlapchange}
\end{align}
The first of these ensures that $\norm{\alpha}^2\vecnorm{\psinot}$ is conserved; since we require $\alpha$ to be a constant, we have that $\vecnorm{\psinot}$ is conserved, as is implied directly by the compact GP form~(\ref{eq:GPcompact}).  Moreover, our formalism assumes that $\ket{\psinot_1}$ and $\ket{\psinot_2}$ remain normalized to unity, implying that any temporal changes in $|\alpha_1|^2\vecnorm{\psinot_1}$ and $|\alpha_2|^2\vecnorm{\psinot_2}$ are incorporated into the magnitudes $\norm{\alpha_1}^2$ and $\norm{\alpha_2}^2$; this simplifies Eqs.~(\ref{eq:normchange}) and~(\ref{eq:overlapchange}) to
\begin{align}\label{eq:normchange2}
\di{}{t}\big(|\alpha_1|^2\big)
&=-\di{}{t}\big(|\alpha_2|^2\big)
=2\,{\rm Im}\big(\omega_{12}\alpha_1^*\alpha_2\braket{\psinot_1}{\psinot_2}\big)\;,\\
i\hbar\di{}{t}\big(\alpha_1^*\alpha_2\braket{\psinot_1}{\psinot_2}\big)
&=\alpha_1^*\alpha_2\bigbra{\psinot_1}\big(H_{\mathrm{gp}}^{(2)}-H_{\mathrm{gp}}^{(1)}\big)\bigket{\psinot_2}
+\hbar\omega_{12}^*\big(\norm{\alpha_1}^2-\norm{\alpha_2}^2\big)\;.
\label{eq:overlapchange2}
\end{align}
Notice also that we can always move any phase changes in $\alpha_1$ and $\alpha_2$ into $\ket{\psinot_1}$ and $\ket{\psinot_2}$; this means that we can always choose $\alpha_1$ and $\alpha_2$ to be real.

If the internal levels are eigenstates of the single-particle Hamiltonian, there is no single-particle coupling of the internal levels, i.e., $\omega_{12}=\omega_{21}^*=0$.  One often uses transient, strong coupling of the internal levels to induce transitions between the internal levels.  This occurs on timescales much shorter than that of the nonlinear terms in the GP equation and can be treated separately as a sudden single-particle effect while ignoring the nonlinear terms; the result is a sudden change in $\alpha_1$ and $\alpha_2$ while $\psinot_1(\mathbf x)$ and $\psinot_2(\mathbf x)$ remain unchanged.  Thus the single-particle coupling terms can generally be omitted when analyzing BEC dynamics; we retain them for completeness in our general development of the Bogoliubov Hamiltonian.

Before moving on, however, we note that if $\omega_{12}=\omega_{21}^*=0$, Eq.~(\ref{eq:normchange2}) implies that $\norm{\alpha_1}$ and $\norm{\alpha_2}$ are constant in time.  Since we can move any phase changes in $\alpha_1$ and $\alpha_2$ into $\ket{\psinot_1}$ and $\ket{\psinot_2}$, we can assume that $\alpha_1$ and $\alpha_2$ are constants, which simplifies Eq.~(\ref{eq:overlapchange2}) to an equation for the change in the overlap of $\ket{\psinot_1}$ and $\ket{\psinot_2}$:
\begin{equation}
i\hbar\di{}{t}\braket{\psinot_1}{\psinot_2}
=\bigbra{\psinot_1}\big(H_{\mathrm{gp}}^{(2)}-H_{\mathrm{gp}}^{(1)}\big)\bigket{\psinot_2}\;.
\end{equation}
Furthermore, it is easy to see from Eq.~(\ref{eq:spinorGP}) that under these circumstances, $\ket{\bar\psinot(t)}$ satisfies the compact GP equation:
\begin{equation}\label{eq:GPcompactbar}
\Big(\mathord{-}i\hbar\pa{}{t}+H_{\mathrm{gp}}\,\Big)\ket{\bar\psinot}=0\;,
\quad\mbox{if $\omega_{12}=\omega_{21}^*=0$.}
\end{equation}

The Bogoliubov Hamiltonian governing the dynamics in the interaction picture is given by Eq.~(\ref{eq:Hbog_two_component}).  In $4\times4$ matrix form, we have
\begin{equation}
\mathcal{H}_\mathrm{bog}= \half\,
\COLON\,
\begin{pmatrix}
\,\bra{\uppsi_1} & \bra{\uppsi_2} & \bra{\uppsi_1^\dagger} & \bra{\uppsi_2^\dagger}\,
\end{pmatrix}
\mathsp{H}_\mathrm{bog}
\begin{pmatrix}
\ket{\uppsi_1} \\[4pt] \ket{\uppsi_2} \\[4pt]\, \ket{\uppsi_1^\dagger}\\[4pt] \ket{\uppsi_2^\dagger}
\end{pmatrix}
\,\COLON\;,
\end{equation}
where the matrix $\mathsp{H}_\mathrm{bog}$ takes the form
\begin{equation}
\label{bog_matrixform_two_component}
\mathsp{H}_\mathrm{bog} =
\scalebox{.95}{\mbox{$\left(\begin{array}{cc|cc}
H_\mathrm{gp}^{(1)}+g_{1\nsp 1} \norm{\alpha_1}^2\norm{\psinot_1}^2
&\hbar\omega_{12}+g_{1\nsp2}\alpha_1 \alpha_2^*\, \psinot_1 \psinot_2^*
&g_{1\nsp 1}\alpha_1^2\, \psinot_1^2
&g_{1\nsp 2}\alpha_1 \alpha_2\,\psinot_1 \psinot_2\\[3pt]
\hbar\omega_{21}+g_{2\nsp 1}\alpha_1^*\alpha_2\,\psinot_1^*\psinot_2
&H_\mathrm{gp}^{(2)}+g_{22}\norm{\alpha_2}^2 \norm{\psinot_2}^2
&g_{2\nsp1}\alpha_1\alpha_2\,\psinot_1\psinot_2
&g_{22}\alpha_2^2\, \psinot_2^2\\[6pt]
\hline&&&\\[-10pt]
g_{1\nsp1} \big(\alpha_1^*)^2 \big(\psinot_1^*\big)^2
&g_{2\nsp1}\alpha_1^*\alpha_2^*\,\psinot_1^*\psinot_2^*
&H_\mathrm{gp}^{(1)}+g_{1\nsp 1} \norm{\alpha_1}^2 \norm{\psinot_1}^2
&\hbar\omega_{21}+g_{2\nsp1}\alpha_1^*\alpha_2\,\psinot_1^*\psinot_2\\[3pt]
g_{1\nsp 2} \alpha_1^*\alpha_2^*\,\psinot_1^*\psinot_2^*
&g_{22}\big(\alpha_2^*)^2 \big(\psinot_2^*\big)^2
&\hbar\omega_{12}+g_{1\nsp 2}\alpha_1 \alpha_2^*\,\psinot_1\psinot_2^* &H_\mathrm{gp}^{(2)}+g_{22}\norm{\alpha_2}^2 \norm{\psinot_2}^2
\end{array}\right)$}}
\;.
\end{equation}
To get back to the compact spinor notation, we introduce, along with the matrix~(\ref{eq:Hgpmatrix}), two other matrices that operate in the spinor space defined by the hyperfine levels $\ket1$ and $\ket2$:
\begin{equation}\label{eq:definition_matrices_two_component}
\Phi=\frac{1}{\alpha}
\begin{pmatrix}
\alpha_1 \psinot_1&
0\\[3pt]
0 &
\alpha_2 \psinot_2
\end{pmatrix}\;,\qquad
G=\begin{pmatrix}
g_{1\nsp 1} &
g_{1\nsp 2}\\[3pt]
g_{2\nsp 1} &
g_{22}
\end{pmatrix}\;.
\end{equation}
With these matrices, we have
\begin{equation}
\mathsp{H}_\mathrm{bog}=
\begin{pmatrix}
H_\mathrm{gp}+\norm\alpha^2\Phi G \Phi^{\nsp *} &
\alpha^2\Phi G \Phi\\[3pt]
(\alpha^*)^2\Phi^{\nsp *} G \Phi^{\nsp *} &
H_\mathrm{gp}^*+\norm\alpha^2\Phi^{\nsp *} G \Phi
\end{pmatrix}\;.
\end{equation}
Notice that since $\Phi$ is diagonal and $G$ is real and symmetric, $\Phi G\Phi^*$ and $\Phi^*G\Phi$ are both Hermitian, and they are transposes and complex conjugates of one another; $\Phi G\Phi$ and $\Phi^* G\Phi^*$ are both symmetric, and they are complex conjugates and Hermitian conjugates of one another.  Using our total field operator and interpreting the $2\times2$ submatrices as operators in the space of the internal levels, we can write the Bogoliubov Hamiltonian in the suggestive form, identical to that for a single component,
\begin{equation}
\mathcal{H}_\mathrm{bog}=\half\,
\COLON\,
\begin{pmatrix}
\bra{\uppsi} & \bra{\uppsi^\dagger}
\end{pmatrix}
\mathsp{H}_\mathrm{bog}
\begin{pmatrix}
\ket{\uppsi}\\[5pt] \ket{\uppsi^\dagger}
\end{pmatrix}
\,\COLON
\;.
\end{equation}

To eliminate phase diffusion in the condensate mode, we now introduce the auxiliary (nonHermitian) Hamiltonian $\sF$ in exactly the same form it has in the single-component case [cf.~Eq.~(\ref{eq:F})],
\begin{align}
\sF(t)=-\frac{\eta(t)}{2}\, (\mathcal{N}-N)^2
+ \big(\alpha\ssp \a^\dagger_{\smash{\psinot(t)}}+\mathcal{N}_{\perp}(t)-N\big)\mathcal{F}_\perp
+ \big(\alpha^* \a_{\psinot(t)}-N\,\big) \mathcal{F}_\perp^\dagger\;,
\label{eq:F_two_component}
\end{align}
where $\mathcal{N}_{\perp}=\mathcal{N}-\a_{\smash \psinot}^\dagger \a_{\psinot}$.  The coefficient $\eta$ and the operator $\sF_\perp=\sF_\perp^\dagger$ are defined in analogy to the single-component case,
\begin{align}
\eta&=\sum_{\sigma,\tau}\frac{1}{\norm{\alpha}^4}
\int g_{\sigma\tau}\norm{\alpha_\sigma}^2\norm{\alpha_\tau}^2 \norm{\psinot_\sigma}^2\norm{\psinot_\tau}^2
\,\dif\mathbf{x}\\
&=\bigbra{\psinot}\Phi G \Phi^{\nsp*}\bigket{\psinot}
=\bigbra{\psinot^*}\Phi^{\nsp*}G \Phi\bigket{\psinot^*}
=\bigbra{\psinot^*}\Phi^{\nsp*}G \Phi^{\nsp*}\bigket{\psinot}
=\bigbra{\psinot}\Phi G \Phi\bigket{\psinot^*}\;,
\end{align}
and
\begin{align}
\sF_\perp
&=-\frac{1}{\norm\alpha^2}
\int\bigg(
\sum_{\sigma,\tau}g_{\sigma\tau}\norm{\alpha_\sigma}^2\norm{\psinot_\sigma}^2
\alpha_\tau^*\psinot_\tau^*\uppsi_{\tau\perp}
+\mathrm{H.c.}\bigg)\,\dif\mathbf x\\[2pt]
&=-\alpha^*\bigbra{\psinot}\Phi G\Phi^{\nsp*}Q\bigket{\uppsi}
-\alpha\bigbra{\psinot}\Phi G \Phi Q^*\bigket{\uppsi^\dagger}
=-\alpha^*\bigbra{\uppsi^\dagger}Q^*\Phi^{\nsp*}G\Phi^{\nsp*}\bigket{\psinot}
-\alpha\bigbra{\uppsi}Q\Phi G \Phi^{\nsp*}\bigket{\psinot}
\;,
\end{align}
where $Q\ket{\uppsi}=\ket{\uppsi_\perp}$ is the total field operator with the condensate mode excluded [see Eq.~(\ref{eq:psiperptwo})].  As in the single-component case, $\eta$ is of order $1/N$ and $\sF_\perp$ is of order $1/N^{1/2}$.  The argument that the auxiliary Hamiltonian $\sF(t)$ does not change the evolution in the $N$-particle sector, as long as the condensate mode stays in a coherent state with amplitude $\alpha$, is the same as that given in the single-component case in Sec.~\ref{subsec:bog}.

The transition to the interaction picture goes exactly as in the single-component case, yielding Eqs.~(\ref{eq:FintBog}) and (\ref{eq:FintBog2}) at Bogoliubov order $N^0$.  Dropping the $c$-number term from that result, we find the analog of Eq.~(\ref{eq:Fint2}):
\begin{align}
\sF_\mathrm{int}
&=-\frac{1}{2}\,\Big(\,
2\norm{\alpha}^2\,\bigbra{\uppsi}P\Phi G\Phi^*P\bigket{\uppsi}
+(\alpha^*)^2\,\bigbra{\uppsi^\dagger}P^*\Phi^*G\Phi^*P\bigket{\uppsi}
+\alpha^2\,\bigbra{\uppsi}P\Phi G\Phi P^*\bigket{\uppsi^\dagger}
\Big)\nonumber\\[2pt]
&\hspace{2em}-\Big(
\norm{\alpha}^2\,\bigbra{\uppsi}P\Phi G\Phi^*Q\bigket{\uppsi}
+\norm{\alpha}^2\,\bigbra{\uppsi}Q\Phi G\Phi^*P\bigket{\uppsi}\nonumber\\
&\qquad\qquad+(\alpha^*)^2\,\bigbra{\uppsi^\dagger}Q^*\Phi^*G\Phi^*P\bigket{\uppsi}
+\alpha^2\,\bigbra{\uppsi}P\Phi G\Phi Q^*\bigket{\uppsi^\dagger}\Big)\;.
\label{eq:Finttwocomp}
\end{align}
Translating this result to matrix form of symplectic structure, we have
\begin{equation}
\sF_\mathrm{int}(t)
=\frac{1}{2}
\,\COLON
\begin{pmatrix}
\bra{\uppsi} & \bra{\uppsi^\dagger}
\end{pmatrix}
\mathsp{F}_{\mathrm{int}}(t)
\begin{pmatrix}
\ket{\uppsi}\\[5pt] \ket{\uppsi^\dagger}
\end{pmatrix}\COLON\;,
\end{equation}
where
\begin{align}
\mathsp{F}_{\mathrm{int}}=
\begin{pmatrix}
|\alpha|^2\big(Q\Phi G\Phi^{\nsp*}Q-\Phi G\Phi^{\nsp*}\big)
&\alpha^2\big(Q\Phi G\Phi Q^*-\Phi G\Phi\big)\\[6pt]
(\alpha^*)^2\big(Q^*\Phi^{\nsp*}G\Phi^{\nsp*} Q-\Phi^{\nsp*}G\Phi^{\nsp*}\big)
&|\alpha|^2\big(Q^*\Phi^{\nsp*}G\Phi Q^*-\Phi^{\nsp*}G\Phi\big)
\end{pmatrix}\;.
\end{align}

The number-conserving Bogoliubov Hamiltonian matrix assumes the form
\begin{equation}\label{eq:Hbogtwofinal}
\mathcal{H}_\mathrm{ncb}=\half\,
\COLON\,
\begin{pmatrix}
\bra{\uppsi} & \bra{\uppsi^\dagger}
\end{pmatrix}
\mathsp{H}_\mathrm{ncb}
\begin{pmatrix}
\ket{\uppsi}\\[5pt] \ket{\uppsi^\dagger}
\end{pmatrix}
\,\COLON
\;,
\end{equation}
with
\begin{align}\label{eq:Hbogtwofinalmatrix}
\mathsp{H}_\mathrm{ncb}
=\mathsp{H}_{\mathrm{bog}}+\mathsp{F}_{\mathrm{int}}
=\begin{pmatrix}
H_\mathrm{gp}+\norm\alpha^2Q\ssp \Phi G \Phi^{\nsp *} Q &
\alpha^2Q\ssp \Phi G \Phi\ssp Q^*\\[3pt]
(\alpha^*)^2Q^* \Phi^{\nsp *} G \Phi^{\nsp *} Q &
H_\mathrm{gp}^*+\norm\alpha^2 Q^*\Phi^{\nsp *} G \Phi\ssp Q^*
\end{pmatrix}\;.
\end{align}

Using the same strategy as in the single-component case, we have derived the number-conserving Bogoliubov Hamiltonian~(\ref{eq:Hbogtwofinalmatrix}) that governs the dynamics of a two-component BEC in the interaction picture.  This Hamiltonian has the same form as the Hamiltonian~(\ref{eq:Hncb_single_component}) that applies in the single-component case; the difference is that here $\Phi$, $G$, $Q$, and $H_\mathrm{gp}$ are themselves matrices.  Our result conforms with Eq.~(3.17) in~\cite{sorensen_bogoliubov_2002}, but in a more compact form.  This compactness is a major advantage in generalizing to the multi-component case.

As in the single-component case, it is useful to divide the number-conserving Bogoliubov Hamiltonian~(\ref{Hncb_matrixform}) into its two natural parts, $\mathcal{H}_\mathrm{ncb}=\mathcal{H}_\mathrm{gp}+\mathcal{\Hrem}$.  The GP part is
\begin{align}\label{eq:Hgpcal2}
\begin{split}
\mathcal{H}_\mathrm{gp}
&=\half\,\COLON
\begin{pmatrix}
\bra{\uppsi} & \bra{\uppsi^\dagger}
\end{pmatrix}\!
\mathsp{H}_\mathrm{gp}
\begin{pmatrix}
\ket{\uppsi}\\[5pt] \ket{\uppsi^\dagger}
\end{pmatrix}\COLON\\[5pt]
&=\bra{\uppsi}H_\mathrm{gp}\ket\uppsi
=\a_{\smash \psinot}^\dagger\a_\psinot\bra{\psinot}H_{\mathrm{gp}}\ket\psinot
+\a_{\smash \psinot}^\dagger\bra\psinot H_{\mathrm{gp}}\ket{\uppsi_\perp}
+\bra{\uppsi_\perp}H_{\mathrm{gp}}\ket\psinot\a_\psinot
+\bra{\uppsi_\perp}H_\mathrm{gp}\ket{\uppsi_\perp}\;,
\end{split}
\end{align}
where
\begin{equation}
\mathsp{H}_\mathrm{gp}
=\begin{pmatrix}
H_\mathrm{gp} & 0\\[6pt]
0 & H_\mathrm{gp}^*
\end{pmatrix}\;.
\end{equation}
The additional coupling of the orthogonal modes, coming from two-body scattering, is
\begin{equation}
\label{eq:Kcal2}
\mathcal{\Hrem}
=\half\,\COLON
\begin{pmatrix}
\bra{\uppsi} & \bra{\uppsi^\dagger}
\end{pmatrix}\!
\mathsp{\Hrem}
\begin{pmatrix}
\ket{\uppsi}\\[5pt] \ket{\uppsi^\dagger}
\end{pmatrix}\COLON
=\half\,\COLON
\begin{pmatrix}
\bra{\uppsi_\perp} & \bra{\uppsi_\perp^\dagger}
\end{pmatrix}\!
\mathsp{\Hrem}
\begin{pmatrix}
\ket{\uppsi_\perp}\\[5pt] \ket{\uppsi_\perp^\dagger}
\end{pmatrix}\COLON\;,
\end{equation}
where
\begin{align}
\label{eq:Ksp2}
\mathsp{\Hrem}
&=\begin{pmatrix}
\norm\alpha^2Q\ssp \Phi G \Phi^{\nsp *} Q & \alpha^2Q\ssp \Phi G \Phi\ssp Q^*\\[4pt]
(\alpha^*)^2Q^* \Phi^{\nsp *} G \Phi^{\nsp *} Q & \norm\alpha^2 Q^*\Phi^{\nsp *} G \Phi\ssp Q^*
\end{pmatrix}
=
\begin{pmatrix}
Q&0\\[4pt]0&Q^*
\end{pmatrix}
\begin{pmatrix}
\norm\alpha^2\ssp \Phi G \Phi^{\nsp *} & \alpha^2\ssp \Phi G \Phi\\[4pt]
(\alpha^*)^2\Phi^{\nsp*} G \Phi^{\nsp*} & \norm\alpha^2\Phi^{\nsp *} G \Phi
\end{pmatrix}
\begin{pmatrix}
Q&0\\[4pt]0&Q^*
\end{pmatrix}\;.
\end{align}

The demonstration that at Bogoliubov order, if the condensate mode begins in a coherent state, it remains in a coherent state can be repeated word for word from the single-component case considered at the beginning of Sec.~\ref{subsec:bogdynamics}, and this shows that the auxiliary Hamiltonian $\sF(t)$ of Eq.~(\ref{eq:F_two_component}) does not change the evolution in the $N$-particle sector.

The Schr\"odinger-picture dynamics can be developed using exactly the same approach and symbology we used in the single-component case, beginning with Eq.~(\ref{eq:Revolution}) and running through the end of Sec.~\ref{subsec:bogdynamics}.  In particular, the Schr\"odinger-picture evolution operator can be written as in Eq.~(\ref{eq:URV}).  The operator $\sR(t)$ evolves according to the GP part of the number-conserving Bogoliubov Hamiltonian, as in Eq.~(\ref{eq:Revolution}), and has the explicit form~(\ref{eq:Rexplicit}).  The operator $\mathcal{V}(t)$ obeys the evolution equation~(\ref{eq:Vevolution}), where the coupling Hamiltonian~$\tilde{\mathcal{\Hrem}}(t)$ is defined by Eq.~(\ref{eq:RGR}).  The matrix~(\ref{eq:tildeK}) generalizes according to Eq.~(\ref{eq:Ksp2}):
\begin{align}
\begin{split}
\tilde{\mathsp{\Hrem}}(t)
&=
\begin{pmatrix}
Q(0)&0\\[4pt]0&Q^*(0)
\end{pmatrix}
\begin{pmatrix}
U_\mathrm{gp}^\dagger(t)&0\\[2pt]0&U_\mathrm{gp}^T(t)
\end{pmatrix}\\[2pt]
&\qquad\times
\begin{pmatrix}
\norm\alpha^2\ssp \Phi(t) G \Phi^{\nsp *}(t) & \alpha^2\ssp \Phi(t) G \Phi(t)\\[4pt]
(\alpha^*)^2\Phi^{\nsp*}(t) G \Phi^{\nsp*}(t) & \norm\alpha^2\Phi^{\nsp *}(t) G \Phi(t)
\end{pmatrix}
\begin{pmatrix}
U_\mathrm{gp}(t)&0\\[2pt]0&U^*_\mathrm{gp}(t)
\end{pmatrix}
\begin{pmatrix}
Q(0)&0\\[4pt]0&Q^*(0)
\end{pmatrix}
\;.
\end{split}
\label{eq:tildeK2}
\end{align}
In terms of a complete set of single-particle states orthogonal to the condensate mode, the coupling Hamiltonian takes on the explicit form,
\begin{align}\label{eq:tildeK2two}
\begin{split}
\tilde{\mathcal{\Hrem}}(t)
&=\norm{\alpha}^2\sum_{j,k\ge1}
\a_j^\dagger\a_k
\bigbra{\psinotnot_j(t)}\Phi(t)G\Phi^*(t)\bigket{\psinotnot_k(t)}\\
&\quad+\frac12\sum_{j,k\ge1}\Big(
(\alpha^*)^2\a_j\a_k\bigbra{\psinotnot_j^*(t)}\Phi^*(t)G\Phi^*(t)\bigket{\psinotnot_k(t)}
+\alpha^2\a_j^\dagger\a_k^\dagger\bigbra{\psinotnot_j(t)}\Phi(t)G\Phi(t)\bigket{\psinotnot^*_k(t)}\Big)\;.
\end{split}
\end{align}

\subsection{Spin squeezing in the Bogoliubov approximation}
\label{subsec:spinsqueeze}

Despite the formal similarity of the single- and two-component cases, there is an important difference, which involves the special orthogonal mode~$\ket{\bar\psinot(t)}$ of Eq.~(\ref{eq:barpsinot}).  In the case of most interest, when there is no single-particle coupling between the internal levels, i.e. $\omega_{12}=\omega_{21}^*=0$, which we specialize to throughout this subsection, we can assume, as we discussed in Sec.~\ref{subsec:twocompbog}, that $\alpha_1$ and $\alpha_2$ are constants in time, and thus the mode $\ket{\bar\psinot}$ of Eq.~(\ref{eq:barpsinot}) satisfies the GP equation~(\ref{eq:GPcompactbar}).  This allows us to make $\ket{\bar\psinot(t)}$ one of the time-dependent modes orthogonal to the condensate mode~$\ket{\psinotnot_0(t)}=\ket{\psinot(t)}$; it is convenient to choose $\ket{\psinotnot_1(t)}=\ket{\bar\psinot(t)}$.  If we further neglect the coupling of $\ket{\bar\psinot}$ to the other orthogonal modes, the coupling Hamiltonian~(\ref{eq:tildeK2two}) reduces to
\begin{align}\label{eq:tildeKbaronly}
\tilde{\mathcal{\Hrem}}(t)
&=\norm{\alpha}^2\ssp\bar\eta(t)\ssp\a_1^\dagger\a_1
+\frac12\ssp\bar\eta(t)\ssp\Big(
\alpha^2(\a_1^\dagger)^2\ssp e^{2\ssp i\ssp\theta}
+(\alpha^*)^2\a_1^2\ssp e^{-2\ssp i\ssp\theta}
\Big)\\
&=\frac12\ssp\norm{\alpha}^2\ssp\bar\eta(t)
\big(\a_1^\dagger\ssp e^{i\mu}+\a_1\ssp e^{-i\mu}\big)^2
-\frac12\ssp\norm{\alpha}^2\ssp\bar\eta(t)\;,
\label{eq:tildeKbaronlytwo}
\end{align}
where we recall that $\a_1=\a_1(0)=\bar a_{\psinot(0)}$, and where we introduce the coupling parameter
\begin{align}\label{eq:etabar}
\begin{split}
\bar\eta(t)
&=\bra{\bar\psinot(t)}\Phi(t)G\Phi^{\nsp *}(t)\ket{\bar\psinot(t)}
=\bra{\bar\psinot^*(t)}\Phi^{\nsp *}(t) G \Phi(t)\ket{\bar\psinot^*(t)}\\[4pt]
&=e^{-2i\theta}\, \bra{\bar\psinot(t)}\Phi(t) G \Phi(t)\ket{\bar\psinot^*(t)}
=e^{2i\theta}\,\bra{\bar{\psinot}^*(t)}\Phi^*(t) G \Phi^*(t)\ket{\bar\psinot(t)}
\end{split}\\
&=\frac{\norm{\alpha_1}^2\ssp \norm{\alpha_2}^2}{\norm{\alpha}^4}
\int \Big(g_{1\nsp 1}\ssp  \norm{\psinot_1}^4+g_{22}\ssp  \norm{\psinot_2}^4-2\ssp g_{1 2}\ssp  \norm{\psinot_1}^2\ssp \norm{\psinot_2}^2\Big)
\dif \mathbf x\;,
\label{eq:etabartwo}
\end{align}
with $\theta = \arg(\alpha_1\alpha_2/\alpha^2)$ being a constant phase angle.  The form~(\ref{eq:tildeKbaronlytwo}), where we let $\alpha e^{i\theta}=\norm{\alpha}e^{i\mu}$, is the Hamiltonian of a free particle with momentum quadrature $(\a_1e^{-i\mu}+\a_1^\dagger e^{i\mu})/\sqrt2$ and a variable mass; this Hamiltonian produces shearing and squeezing in the direction of the position quadrature at a variable rate given by $2\norm{\alpha}^2\bar\eta(t)$.

We can solve for the Heisenberg-picture evolution of $\a_1$,
\begin{equation}
\mathcal{V}^\dagger(t)\a_1\mathcal{V}(t)
=\a_1\big[1-i\norm{\alpha}^2\bar\xi(t)\big]
-i\a_1^\dagger\norm{\alpha}^2 e^{2i\mu}\bar\xi(t)
=e^{2i\nu(t)\a_1^\dagger\a_1}S^\dagger\big(\zeta(t)\big)
\a_1
S\big(\zeta(t)\big) e^{-2i\nu(t)\a_1^\dagger\a_1}\;.
\end{equation}
In the first form,
\begin{equation}
\bar\xi(t)=\frac{1}{\hbar}\int_0^t\bar\eta(t')\,dt'
\end{equation}
is the dimensionless integral of the coupling parameter.  In the second form,
\begin{equation}
S(\zeta)=\exp\bigg(\frac12\big[\zeta^*\a_1^2-\zeta(\a_1^\dagger)^2\big]\bigg)
\end{equation}
is the single-mode squeeze operator~\cite{perelomov_generalized_1977, hollenhorst_quantum_1979,Caves1985a,Schumaker1985a,Schumaker1986a}, with the complex squeezing parameter given by $\zeta(t)=i\gamma(t)e^{2i\mu}e^{-2i\nu(t)}$, where
\begin{equation}
\sinh \gamma(t)=\norm{\alpha}^2\bar\xi(t)=\tan[2\nu(t)]\;.
\end{equation}

These results determine the evolution operator $\mathcal{V}(t)$ up to a phase.  Irrelevant though it is, the phase can be determined by integrating directly the evolution equation for $\mathcal{V}(t)$ or by considering the vacuum expectation value of $\mathcal{V}(t)$, with the result that
\begin{equation}\label{eq:decomposition}
\mathcal{V}(t)=e^{i\upsilon(t)}S\big(\zeta(t)\big)e^{-2i\nu(t)\a_1^\dagger\a_1}\;,
\end{equation}
where $\upsilon(t)=\frac12\norm{\alpha}^2\bar\xi(t)-
\frac12\tan^{-1}[\norm{\alpha}^2\bar\xi(t)]$.

Suppose now that the orthogonal mode $\ket{\bar\psinot}$ begins in vacuum, so that the initial extended catalytic state is $\ket{\psibf_\mathrm{ecs}(0)}=\ket{\alpha,\psinot(0)}_0\otimes\ket{\mathrm{vac},\bar\psinot(0)}_1$.  Then the extended catalytic state at time~$t$~is
\begin{equation}
\ket{\psibf_\mathrm{ecs}(t)}
=\sR(t)\mathcal{V}(t)\ket{\psibf_\mathrm{ecs}(0)}
=e^{i\upsilon(t)}\ket{\alpha,\psinot(t)}_0
\otimes\sRperp(t)S\big(\zeta(t)\big)\ket{\mathrm{vac},\bar\psinot(0)}_1\;.
\end{equation}
where
\begin{equation}
\sRperp(t)=\sum_{n=0}^\infty\ket{n,\bar\psinot(t)}_1\bra{n,\bar\psinot(0)}_1\;.
\end{equation}
The physical state at time~$t$ follows from projecting onto the $N$-particle sector, as specified by Eq.~(\ref{eq:projection}), giving
\begin{equation}\label{eq:Bogsqueezing}
\ket{\psibf_N(t)}
=\sum_{M=0}^N\ket{N-M,\psinot(t)}_0\otimes\ket{M,\bar\psinot(t)}_1
\frac{1}{\alpha^M}\sqrt{\frac{N!}{(N-M)!}}
\,{}_1\bra{M,\bar\psinot(0)}S\big(\zeta(t)\big)\ket{\mathrm{vac},\bar\psinot(0)}_1
\end{equation}
The matrix elements of the squeeze operator can be evaluated explicitly (only even values of $M$ have nonzero matrix elements), but we do not bother with that here, noting instead that the Bogliubov approximation requires that these matrix elements be small for $M\ge2$.  A more quantitative statement is that these matrix elements must be $\alt1/\sqrt N$ for $M\ge2$, which translates to small squeezing with $\norm{\gamma(t)}\alt1/\sqrt N$ or $\norm{\bar\xi(t)}\alt1/N^{3/2}$.  In practice, since the elements of $G$ are nearly equal (they typically differ only by small differences in scattering length for the hyperfine levels), the smallness of $\norm{\bar\xi(t)}$ or $\norm{\bar\eta(t)}$ is governed by the difference in the probability densities for the two internal levels, $\norm{\psinot_1({\mathbf x})}^2$ and $\norm{\psinot_2({\mathbf x})}^2$; roughly speaking, the Bogliobov approximation requires that the two hyperfine levels not  be separated spatially.

If $\norm{\bar\xi(t)}$ becomes too large, perhaps due to spatial separation of the internal levels, one can do a better job by returning to the Hamiltonian~(\ref{eq:tildeKbaronly}) and recalling that  it arises, in the Bogoliubov approximation, from replacing $\a_\psinot$ and $\a_{\smash\psinot}^\dagger$ by $\alpha$ and $\alpha^{\nsp*}$ in the original Schr\"odinger-picture Hamiltonian.  Restoring, in normal order, the creation and annihilation operators for the condensate mode to $\tilde{\mathcal{\Hrem}}(t)$ gives a Kerr-like interaction between the condensate mode~$\ket{\psinot}$ and the orthogonal mode~$\ket{\bar\psinot}$,
\begin{align}\label{eq:Kerra}
\tilde{\mathcal{\Hrem}}(t)
&=\frac{1}{2}\bar\eta(t)\,
\Big(\ssp 2\ssp\a_0^\dagger\a_1^\dagger\ssp\a_0\ssp\a_1
+e^{2\ssp i\ssp\theta}\ssp(\a_1^\dagger)^2\ssp\a_0^2\ssp
+e^{-2\ssp i\ssp\theta}\ssp(\a_0^\dagger)^2\ssp\a_1^2\ssp\Big)\\
&=\frac{1}{2}\bar\eta(t)\,
\big(\ssp e^{i\ssp\theta}\ssp\a_1^\dagger\ssp\a_0
+e^{-i\ssp\theta}\ssp\a_0^\dagger\ssp\a_1\ssp\big)^2
-\frac{1}{2}\bar\eta(t)
\big(\a_0^\dagger\a_0+\a_1^\dagger\a_1\big)\;,
\label{eq:Kerrb}
\end{align}
where $\a_0=\a_{\psinot(0)}$.  The first term in Eq.~(\ref{eq:Kerra}) comes from scattering of $\ket\psinot$- and $\ket{\bar\psinot}$-particles off one another, the second term from scattering of two $\ket\psinot$-particles into the $\ket{\bar\psinot}$-mode, and the last term from scattering of two $\ket{\bar\psinot}$-particles into the $\ket\psinot$-mode.

The Hamiltonian $\tilde{\mathcal{\Hrem}}(t)$ conserves the total particle number $\mathcal{N}=\a_0^\dagger\a_0+\a_1^\dagger\a_1$.  Thus, if the initial state is in the $N$-particle sector of the two modes, it stays there, and we can omit the projection onto the $N$-particle sector that is involved in our use of an extended catalytic state.  Moreover, the last term in Eq.~(\ref{eq:Kerrb}) is proportional to $\mathcal{N}$ and thus becomes the c-number $N\bar\eta(t)/2$; this term only introduces an irrelevant overall phase, so we can neglect it.  Finally, we let $\alpha$, $\alpha_1$, and $\alpha_2$ be real and positive, which makes $\theta=0$, thus leaving us with the Hamiltonian
\begin{align}\label{eq:one_axis_twist}
\tilde{\mathcal{\Hrem}}_\textrm{ss}(t)
= 2\ssp \bar\eta(t)\mathcal{J}_x^2 \;,
\end{align}
where
\begin{align}
\mathcal{J}_x
\equiv\frac{1}{2}\, \big(\ssp \a_0^\dagger \a_1+\a_1^\dagger \a_0\ssp \big)
\end{align}
is the $x$-component of the Schwinger pseudo-spin of the two modes.  The other two Schwinger operators are $\mathcal{J}_y=-i\big(\ssp\a_0^\dagger\a_1-\a_1^\dagger\a_0\ssp \big)/2$ and $\mathcal{J}_z=\big(\ssp\a_0^\dagger\a_0-\a_1^\dagger\a_1\ssp\big)/2$.

At this point the model has been reduced to two modes, both of which participate in the dynamics.  The single-particle states for the two modes, $\ket{\psinot(t)}$ and $\ket{\bar\psinot(t)}$, change in time according to the GP equations~(\ref{eq:GPcompact}) and~(\ref{eq:GPcompactbar}), both of which are expressions of the coupled GP equations~(\ref{eq:GP_two_component}) (with $\omega_{12}=\omega_{21}^*=0$).  The state of the two modes changes according to the evolution operator $\mathcal{V}(t)$ of Eq.~(\ref{eq:Vevolution}), where one uses the two-mode Hamiltonian $\tilde{\mathcal{\Hrem}}_\textrm{ss}(t)$, which has fixed creation and annihilation operators; this evolution is followed by application of the operator $\sR(t)$ of Eq.~(\ref{eq:Rexplicit}), which translates the two-mode state to the modes that apply at time~$t$.

This model ignores the coupling of the two dominant modes to the other orthogonal modes, but its chief problem lies in an inconsistency between the GP equations~(\ref{eq:GP_two_component}) and the evolution under the Hamiltonian~(\ref{eq:one_axis_twist}).  The GP equations contain the quantities $\norm{\alpha_1}^2$ and $\norm{\alpha_2}^2$, which can be interpreted as mean particle numbers for the two internal levels, $N_1=\norm{\alpha_1}^2$ and $N_2=\norm{\alpha_2}^2$; the internal levels initially have number uncertainties of order $\sqrt{N_1}$ and $\sqrt{N_2}$, which are small compared to the mean particle numbers.  As discussed previously, the GP equations~(\ref{eq:GP_two_component}) leave $\norm{\alpha_1}^2$ and $\norm{\alpha_2}^2$ unchanged when $\omega_{12}=\omega_{21}^*=0$.  The inconsistency arises because the number operators for the two internal levels are generally not conserved by $\tilde{\mathcal{\Hrem}}_\textrm{ss}(t)$.  For the model to be consistent, the number operators for internal levels, $\mathcal{N}_1=b_1^\dagger b_1$ and $\mathcal{N}_2=b_2^\dagger b_2$, should be conserved or nearly so, so that the mean particle numbers don't change and the uncertainties remain small.

Letting $\alpha_1/\alpha=\cos(\varphi/2)$ and $\alpha_2/\alpha=\sin(\varphi/2)$, we have, from Eqs.~(\ref{eq:ab}) and~(\ref{eq:abarb}), $b_1=b_{1,\psinot_1}=\a_0\cos(\varphi/2)+\a_1\sin(\varphi/2)$ and $b_2=b_{2,\psinot_2}=\a_0\sin(\varphi/2)-\a_1\cos(\varphi/2)$ and thus
\begin{align}
\begin{split}
\mathcal{N}_1&=b_1^\dagger b_1
=\frac12\mathcal{N}+\mathcal{J}_z\cos\varphi+\mathcal{J}_x\sin\varphi\;,\\
\mathcal{N}_2&=b_2^\dagger b_2
=\frac12\mathcal{N}-\mathcal{J}_z\cos\varphi-\mathcal{J}_x\sin\varphi\;.
\end{split}
\end{align}
Thus the condition for the model to be consistent is that the condensate mode be an equal superposition of the two internal levels, i.e., $\varphi=\pi/2$, or nearly so.  The combination of the GP equations~(\ref{eq:GP_two_component}) for evolving the spatial mode functions and the Hamiltonian~(\ref{eq:one_axis_twist}) to evolve the two-mode state in the case of an equal superposition of the internal levels is called the two-component formalism (or two-mode approximation)~\cite{sorensen_bogoliubov_2002,Poulsen2001a,poulsen_bose-einstein_2002}.  The two-mode approximation is more robust than the Bogoliubov-approximation squeezing results summarized in Eq.~(\ref{eq:Bogsqueezing}).  In the context of condensates isolated in fairly well separated trapping potentials, very recent work has analyzed the effect of including four macroscopically excited modes, two in each well~\cite{gillet_tunneling_2014}.

The $\mathcal{J}_x^2$ term in $\tilde{\mathcal{\Hrem}}_\textrm{ss}$ is the so-called one-axis-twisting Hamiltonian~\cite{kitagawa_squeezed_1993}; it induces spin squeezing in states that are initially maximally polarized along the spin $z$ axis, as is the case for an initial state that has all $N$ particles in the condensate mode $\ket\psinot$.  The one-axis-twisting Hamiltonian is widely used to generate spin squeezing in BECs~\cite{esteve_squeezing_2008, grond_optimizing_2009, riedel_atom-chip-based_2010, gross_nonlinear_2010}.  Of particular relevance to our formulation are analyses of the interplay of spatial and spin dynamics~\cite{sinatra_binary_2000, sorensen_many-particle_2001, li_spin_2009}; in addition, Sinatra \emph{et al.}~\cite{sinatra_limit_2011} showed that the amount of squeezing is bounded from above by the initial noncondensed fraction at finite temperature.

\section{Summary and Conclusion}
\label{sec:summary}

In this paper we develop a new framework for deriving the number-conserving Bogoliubov approximation for a dilute-gas BEC.  Our approach begins by introducing the extended catalytic state~(\ref{eq:catalytic_state}), a coherent state for the condensate mode and an arbitrary state for the modes orthogonal to the condensate mode.  The physical state with exactly $N$ particles is retrieved from the extended catalytic state by projecting into the $N$-particle sector, as in Eq.~(\ref{eq:psiN}).  To formulate the Bogoliubov approximation, we introduce the time-dependent interaction picture~(\ref{eq:intpic}) in which the condensate mode is displaced to the vacuum. The field operators are thus of order $N^0$, and we can organize the BEC Hamiltonian in powers of ${N}^{-1/2}$.  Requiring the terms of order ${N}^{1/2}$ to vanish yields the Gross-Pitaevskii equation~(\ref{eq:GPE}).  Going to the next order, $N^0$, gives the conventional Bogoliubov Hamiltonian of Eqs.~(\ref{eq:Hbog_matrixform}) and (\ref{eq:Hbog_matrix}).  Introducing the auxiliary Hamiltonian~(\ref{eq:F}) removes the unwanted phase diffusion from the conventional Bogoliubov Hamiltonian without affecting the physical state in the $N$-particle sector (to order $N^0$).  The result is the number-conserving Bogoliubov Hamiltonian of Eqs.~(\ref{Hncb_matrixform}) and (\ref{eq:Hncb_single_component}).  Analysis of the dynamics under the number-conserving Bogoliubov Hamiltonian shows that its GP part~(\ref{eq:Hgpcal}) evolves the single-particle, spatial mode structure forward in time, and the remaining part~(\ref{eq:Kcal}) evolves the state of these modes.  The result is the particularly simple form~(\ref{eq:URV}) for the Schr\"odinger-picture evolution at Bogoliubov order.

In Sec.~\ref{sec:twocomp} we extend our approach to BECs with two internal levels.  This turns out to be largely a matter of using a spinor notation that puts the derivation into a form that mimics the single-component derivation of Sec.~\ref{sec:ecsbog}.  Because of this formal similarity, generalization to multiple hyperfine levels would be straightforward.  In Sec.~\ref{subsec:spinsqueeze} we specialize the Bogoliubov approximation to the two dominant modes, the condensate mode and the mode orthogonal to it in the same two-dimensional subspace of the internal levels.  We discuss how to generalize beyond the Bogoliubov approximation to the two-mode approximation for these two dominant modes, thus allowing a treatment of the spin squeezing of these two modes.

The number-conserving Bogoliubov Hamiltonian we find is identical to the one derived originally by Castin and Dum~\cite{castin_low-temperature_1998}.  The chief difference between our approach and that of Castin and Dum is that they worked in the Heisenberg picture, whereas our derivation is carried out in the Schr\"odinger picture and a closely allied interaction picture.  There are several reasons for presenting a new framework for a derivation of the same result.  The first is the modest one that the new derivation might highlight assumptions from a different perspective and ease the way forward on different, but related problems.  Indeed, in our approach, the Bogoliubov excitations are easily seen as excitations on top of a pure condensate mode that evolves according to the GP equation.  The method we use for handling the phase diffusion that arises from assigning a phase to the condensate mode comes directly from re-asserting the number conservation that applies to a lossless BEC. \ A second reason is that working in the Schr\"odinger picture allows us to separate cleanly, within the Bogoliubov approximation, the evolution of the spatial mode structure from the evolution of the state of these modes.  A central problem of dilute-gas BEC theory and of many other problems in many-body physics is how to do this separation appropriately, and our analysis can be instructive in how to formulate this separation.   A third reason is that the Schr\"odinger picture allows us to identify the entanglement between the condensate mode and the orthogonal modes that are excited by the Bogoliubov Hamiltonian.  In our approach, this entanglement arises when the extended catalytic state is projected into the $N$-particle sector to obtain the physical state of the BEC. \ The entanglement is something that can be analyzed easily in our Schr\"odinger-picture formulation, whereas even the proper formulation of entanglement is difficult in the Heisenberg picture.

These three reasons motivated our work on this topic.  We trust that they justify its presentation to the wider scientific community.

\acknowledgments
The authors thanks A.~B. Tacla for useful and stimulating conversations. This work was supported in part by National Science Foundation Grants No. PHY-1212445 and No. PHY-1314763 and by Office of Naval Research Grant No. N00014-15-1-2167.

%


\begin{thebibliography}{49}%
\makeatletter
\providecommand \@ifxundefined [1]{%
 \@ifx{#1\undefined}
}%
\providecommand \@ifnum [1]{%
 \ifnum #1\expandafter \@firstoftwo
 \else \expandafter \@secondoftwo
 \fi
}%
\providecommand \@ifx [1]{%
 \ifx #1\expandafter \@firstoftwo
 \else \expandafter \@secondoftwo
 \fi
}%
\providecommand \natexlab [1]{#1}%
\providecommand \enquote  [1]{``#1''}%
\providecommand \bibnamefont  [1]{#1}%
\providecommand \bibfnamefont [1]{#1}%
\providecommand \citenamefont [1]{#1}%
\providecommand \href@noop [0]{\@secondoftwo}%
\providecommand \href [0]{\begingroup \@sanitize@url \@href}%
\providecommand \@href[1]{\@@startlink{#1}\@@href}%
\providecommand \@@href[1]{\endgroup#1\@@endlink}%
\providecommand \@sanitize@url [0]{\catcode `\\12\catcode `\$12\catcode
  `\&12\catcode `\#12\catcode `\^12\catcode `\_12\catcode `\%12\relax}%
\providecommand \@@startlink[1]{}%
\providecommand \@@endlink[0]{}%
\providecommand \url  [0]{\begingroup\@sanitize@url \@url }%
\providecommand \@url [1]{\endgroup\@href {#1}{\urlprefix }}%
\providecommand \urlprefix  [0]{URL }%
\providecommand \Eprint [0]{\href }%
\providecommand \doibase [0]{http://dx.doi.org/}%
\providecommand \selectlanguage [0]{\@gobble}%
\providecommand \bibinfo  [0]{\@secondoftwo}%
\providecommand \bibfield  [0]{\@secondoftwo}%
\providecommand \translation [1]{[#1]}%
\providecommand \BibitemOpen [0]{}%
\providecommand \bibitemStop [0]{}%
\providecommand \bibitemNoStop [0]{.\EOS\space}%
\providecommand \EOS [0]{\spacefactor3000\relax}%
\providecommand \BibitemShut  [1]{\csname bibitem#1\endcsname}%
\let\auto@bib@innerbib\@empty
\bibitem [{\citenamefont {Bogoliubov}(1947)}]{bogoliubov_theory_1947}%
  \BibitemOpen
  \bibfield  {author} {\bibinfo {author} {\bibfnamefont {N.~N.}\ \bibnamefont
  {Bogoliubov}},\ }\bibfield  {title} {\emph {\bibinfo {title} {On the theory
  of superfluidity}},\ }\href@noop {} {\bibfield  {journal} {\bibinfo
  {journal} {Journal of Physics (USSR)}\ }\textbf {\bibinfo {volume} {11}},\
  \bibinfo {pages} {23} (\bibinfo {year} {1947})}\BibitemShut {NoStop}%
\bibitem [{\citenamefont {Fetter}(1972)}]{fetter_nonuniform_1972}%
  \BibitemOpen
  \bibfield  {author} {\bibinfo {author} {\bibfnamefont {A.~L.}\ \bibnamefont
  {Fetter}},\ }\bibfield  {title} {\emph {\bibinfo {title} {Nonuniform states
  of an imperfect {Bose} gas}},\ }\href {\doibase 10.1016/0003-4916(72)90330-2}
  {\bibfield  {journal} {\bibinfo  {journal} {Annals of Physics}\ }\textbf
  {\bibinfo {volume} {70}},\ \bibinfo {pages} {67} (\bibinfo {year}
  {1972})}\BibitemShut {NoStop}%
\bibitem [{\citenamefont {Huang}(1987)}]{huang_statistical_1987}%
  \BibitemOpen
  \bibfield  {author} {\bibinfo {author} {\bibfnamefont {K.}~\bibnamefont
  {Huang}},\ }\href@noop {} {\emph {\bibinfo {title} {Statistical Mechanics,
  {\rm 2nd Edition}}}}\ (\bibinfo  {publisher} {Wiley},\ \bibinfo {year}
  {1987})\BibitemShut {NoStop}%
\bibitem [{\citenamefont {Nambu}(1960)}]{nambu_quasi-particles_1960}%
  \BibitemOpen
  \bibfield  {author} {\bibinfo {author} {\bibfnamefont {Y.}~\bibnamefont
  {Nambu}},\ }\bibfield  {title} {\emph {\bibinfo {title} {Quasi-particles and
  gauge invariance in the theory of superconductivity}},\ }\href {\doibase
  10.1103/PhysRev.117.648} {\bibfield  {journal} {\bibinfo  {journal} {Physical
  Review}\ }\textbf {\bibinfo {volume} {117}},\ \bibinfo {pages} {648}
  (\bibinfo {year} {1960})}\BibitemShut {NoStop}%
\bibitem [{\citenamefont {Goldstone}(1961)}]{goldstone_field_1961}%
  \BibitemOpen
  \bibfield  {author} {\bibinfo {author} {\bibfnamefont {J.}~\bibnamefont
  {Goldstone}},\ }\bibfield  {title} {\emph {\bibinfo {title} {Field theories
  with superconductor solutions}},\ }\href {\doibase 10.1007/BF02812722}
  {\bibfield  {journal} {\bibinfo  {journal} {Il Nuovo Cimento}\ }\textbf
  {\bibinfo {volume} {19}},\ \bibinfo {pages} {154} (\bibinfo {year}
  {1961})}\BibitemShut {NoStop}%
\bibitem [{\citenamefont {Girardeau}\ and\ \citenamefont
  {Arnowitt}(1959)}]{girardeau_theory_1959}%
  \BibitemOpen
  \bibfield  {author} {\bibinfo {author} {\bibfnamefont {M.}~\bibnamefont
  {Girardeau}}\ and\ \bibinfo {author} {\bibfnamefont {R.}~\bibnamefont
  {Arnowitt}},\ }\bibfield  {title} {\emph {\bibinfo {title} {Theory of
  many-boson systems: Pair theory}},\ }\href {\doibase 10.1103/PhysRev.113.755}
  {\bibfield  {journal} {\bibinfo  {journal} {Physical Review}\ }\textbf
  {\bibinfo {volume} {113}},\ \bibinfo {pages} {755} (\bibinfo {year}
  {1959})}\BibitemShut {NoStop}%
\bibitem [{\citenamefont
  {Gardiner}(1997)}]{gardiner_particle-number-conserving_1997}%
  \BibitemOpen
  \bibfield  {author} {\bibinfo {author} {\bibfnamefont {C.~W.}\ \bibnamefont
  {Gardiner}},\ }\bibfield  {title} {\emph {\bibinfo {title}
  {Particle-number-conserving {B}ogoliubov method which demonstrates the
  validity of the time-dependent {G}ross-{P}itaevskii equation for a highly
  condensed {B}ose gas}},\ }\href {\doibase 10.1103/PhysRevA.56.1414}
  {\bibfield  {journal} {\bibinfo  {journal} {Physical Review A}\ }\textbf
  {\bibinfo {volume} {56}},\ \bibinfo {pages} {1414} (\bibinfo {year}
  {1997})}\BibitemShut {NoStop}%
\bibitem [{\citenamefont {Gardiner}\ \emph {et~al.}(1997)\citenamefont
  {Gardiner}, \citenamefont {Zoller}, \citenamefont {Ballagh},\ and\
  \citenamefont {Davis}}]{gardiner_kinetics_1997}%
  \BibitemOpen
  \bibfield  {author} {\bibinfo {author} {\bibfnamefont {C.~W.}\ \bibnamefont
  {Gardiner}}, \bibinfo {author} {\bibfnamefont {P.}~\bibnamefont {Zoller}},
  \bibinfo {author} {\bibfnamefont {R.~J.}\ \bibnamefont {Ballagh}}, \ and\
  \bibinfo {author} {\bibfnamefont {M.~J.}\ \bibnamefont {Davis}},\ }\bibfield
  {title} {\emph {\bibinfo {title} {Kinetics of {Bose-Einstein} condensation in
  a trap}},\ }\href {\doibase 10.1103/PhysRevLett.79.1793} {\bibfield
  {journal} {\bibinfo  {journal} {Physical Review Letters}\ }\textbf {\bibinfo
  {volume} {79}},\ \bibinfo {pages} {1793} (\bibinfo {year}
  {1997})}\BibitemShut {NoStop}%
\bibitem [{\citenamefont {Castin}\ and\ \citenamefont
  {Dum}(1997)}]{castin_instability_1997}%
  \BibitemOpen
  \bibfield  {author} {\bibinfo {author} {\bibfnamefont {Y.}~\bibnamefont
  {Castin}}\ and\ \bibinfo {author} {\bibfnamefont {R.}~\bibnamefont {Dum}},\
  }\bibfield  {title} {\emph {\bibinfo {title} {Instability and depletion of an
  excited {Bose-Einstein} condensate in a trap}},\ }\href {\doibase
  10.1103/PhysRevLett.79.3553} {\bibfield  {journal} {\bibinfo  {journal}
  {Physical Review Letters}\ }\textbf {\bibinfo {volume} {79}},\ \bibinfo
  {pages} {3553} (\bibinfo {year} {1997})}\BibitemShut {NoStop}%
\bibitem [{\citenamefont {Castin}\ and\ \citenamefont
  {Dum}(1998)}]{castin_low-temperature_1998}%
  \BibitemOpen
  \bibfield  {author} {\bibinfo {author} {\bibfnamefont {Y.}~\bibnamefont
  {Castin}}\ and\ \bibinfo {author} {\bibfnamefont {R.}~\bibnamefont {Dum}},\
  }\bibfield  {title} {\emph {\bibinfo {title} {Low-temperature
  {B}ose-{E}instein condensates in time-dependent traps: Beyond the
  \mbox{$U$}(1) symmetry-breaking approach}},\ }\href {\doibase
  10.1103/PhysRevA.57.3008} {\bibfield  {journal} {\bibinfo  {journal}
  {Physical Review A}\ }\textbf {\bibinfo {volume} {57}},\ \bibinfo {pages}
  {3008} (\bibinfo {year} {1998})}\BibitemShut {NoStop}%
\bibitem [{\citenamefont {Castin}(2001)}]{castin_bose-einstein_2001}%
  \BibitemOpen
  \bibfield  {author} {\bibinfo {author} {\bibfnamefont {Y.}~\bibnamefont
  {Castin}},\ }\bibfield  {title} {\emph {\bibinfo {title} {{Bose-Einstein}
  condensates in atomic gases: Simple theoretical results}},\ }in\ \href
  {http://link.springer.com/chapter/10.1007/3-540-45338-5_1} {\emph {\bibinfo
  {booktitle} {Coherent Atomic Matter Waves}}},\ \bibinfo {series and number}
  {\bibinfo {series} {Les Houches - Ecole {d'Ete} de Physique Theorique}\
  No.~\bibinfo {number} {72}},\ \bibinfo {editor} {edited by\ \bibinfo {editor}
  {\bibfnamefont {R.}~\bibnamefont {Kaiser}}, \bibinfo {editor} {\bibfnamefont
  {C.}~\bibnamefont {Westbrook}}, \ and\ \bibinfo {editor} {\bibfnamefont
  {F.}~\bibnamefont {David}}}\ (\bibinfo  {publisher} {Springer},\ \bibinfo
  {address} {Berlin},\ \bibinfo {year} {2001})\ pp.\ \bibinfo {pages}
  {1--136}\BibitemShut {NoStop}%
\bibitem [{\citenamefont {S{\o}rensen}(2002)}]{sorensen_bogoliubov_2002}%
  \BibitemOpen
  \bibfield  {author} {\bibinfo {author} {\bibfnamefont {A.~S.}\ \bibnamefont
  {S{\o}rensen}},\ }\bibfield  {title} {\emph {\bibinfo {title} {{Bogoliubov}
  theory of entanglement in a {Bose-Einstein} condensate}},\ }\href {\doibase
  10.1103/PhysRevA.65.043610} {\bibfield  {journal} {\bibinfo  {journal}
  {Physical Review A}\ }\textbf {\bibinfo {volume} {65}},\ \bibinfo {pages}
  {043610} (\bibinfo {year} {2002})}\BibitemShut {NoStop}%
\bibitem [{\citenamefont {Gardiner}\ and\ \citenamefont
  {Morgan}(2007)}]{gardiner_number-conserving_2007}%
  \BibitemOpen
  \bibfield  {author} {\bibinfo {author} {\bibfnamefont {S.~A.}\ \bibnamefont
  {Gardiner}}\ and\ \bibinfo {author} {\bibfnamefont {S.~A.}\ \bibnamefont
  {Morgan}},\ }\bibfield  {title} {\emph {\bibinfo {title} {Number-conserving
  approach to a minimal self-consistent treatment of condensate and
  noncondensate dynamics in a degenerate {Bose} gas}},\ }\href {\doibase
  10.1103/PhysRevA.75.043621} {\bibfield  {journal} {\bibinfo  {journal}
  {Physical Review A}\ }\textbf {\bibinfo {volume} {75}},\ \bibinfo {pages}
  {043621} (\bibinfo {year} {2007})}\BibitemShut {NoStop}%
\bibitem [{\citenamefont {{S. A.~Gardiner}}\ and\ \citenamefont {{T.
  P.~Billam}}(2011)}]{simon_a._gardiner_number-conserving_2011}%
  \BibitemOpen
  \bibfield  {author} {\bibinfo {author} {\bibnamefont {{S. A.~Gardiner}}}\
  and\ \bibinfo {author} {\bibnamefont {{T. P.~Billam}}},\ }\bibfield  {title}
  {\emph {\bibinfo {title} {Number-conserving approaches for atomic
  {B}ose-{E}instein condensates: An overview}},\ }in\ \href
  {http://www.worldscientific.com/doi/abs/10.1142/9781848168121_0008} {\emph
  {\bibinfo {booktitle} {Quantum Gases}}},\ \bibinfo {series} {Cold Atoms},
  Vol.~\bibinfo {volume} {1}\ (\bibinfo  {publisher} {Imperial College Press},\
  \bibinfo {year} {2011})\ pp.\ \bibinfo {pages} {133--145}\BibitemShut
  {NoStop}%
\bibitem [{\citenamefont {Mason}\ and\ \citenamefont
  {Gardiner}(2014)}]{mason_number-conserving_2014}%
  \BibitemOpen
  \bibfield  {author} {\bibinfo {author} {\bibfnamefont {P.}~\bibnamefont
  {Mason}}\ and\ \bibinfo {author} {\bibfnamefont {S.~A.}\ \bibnamefont
  {Gardiner}},\ }\bibfield  {title} {\emph {\bibinfo {title} {Number-conserving
  approaches to $n$-component {B}ose-{E}instein condensates}},\ }\href
  {\doibase 10.1103/PhysRevA.89.043617} {\bibfield  {journal} {\bibinfo
  {journal} {Physical Review A}\ }\textbf {\bibinfo {volume} {89}},\ \bibinfo
  {pages} {043617} (\bibinfo {year} {2014})}\BibitemShut {NoStop}%
\bibitem [{\citenamefont {Steel}\ \emph {et~al.}(1998)\citenamefont {Steel},
  \citenamefont {Olsen}, \citenamefont {Plimak}, \citenamefont {Drummond},
  \citenamefont {Tan}, \citenamefont {Collett}, \citenamefont {Walls},\ and\
  \citenamefont {Graham}}]{steel_dynamical_1998}%
  \BibitemOpen
  \bibfield  {author} {\bibinfo {author} {\bibfnamefont {M.~J.}\ \bibnamefont
  {Steel}}, \bibinfo {author} {\bibfnamefont {M.~K.}\ \bibnamefont {Olsen}},
  \bibinfo {author} {\bibfnamefont {L.~I.}\ \bibnamefont {Plimak}}, \bibinfo
  {author} {\bibfnamefont {P.~D.}\ \bibnamefont {Drummond}}, \bibinfo {author}
  {\bibfnamefont {S.~M.}\ \bibnamefont {Tan}}, \bibinfo {author} {\bibfnamefont
  {M.~J.}\ \bibnamefont {Collett}}, \bibinfo {author} {\bibfnamefont {D.~F.}\
  \bibnamefont {Walls}}, \ and\ \bibinfo {author} {\bibfnamefont
  {R.}~\bibnamefont {Graham}},\ }\bibfield  {title} {\emph {\bibinfo {title}
  {Dynamical quantum noise in trapped {Bose-Einstein} condensates}},\ }\href
  {\doibase 10.1103/PhysRevA.58.4824} {\bibfield  {journal} {\bibinfo
  {journal} {Physical Review A}\ }\textbf {\bibinfo {volume} {58}},\ \bibinfo
  {pages} {4824} (\bibinfo {year} {1998})}\BibitemShut {NoStop}%
\bibitem [{\citenamefont {Sinatra}\ \emph {et~al.}(2002)\citenamefont
  {Sinatra}, \citenamefont {Lobo},\ and\ \citenamefont
  {Castin}}]{sinatra_truncated_2002}%
  \BibitemOpen
  \bibfield  {author} {\bibinfo {author} {\bibfnamefont {A.}~\bibnamefont
  {Sinatra}}, \bibinfo {author} {\bibfnamefont {C.}~\bibnamefont {Lobo}}, \
  and\ \bibinfo {author} {\bibfnamefont {Y.}~\bibnamefont {Castin}},\
  }\bibfield  {title} {\emph {\bibinfo {title} {The truncated {Wigner} method
  for {Bose}-condensed gases: limits of validity and applications}},\ }\href
  {\doibase 10.1088/0953-4075/35/17/301} {\bibfield  {journal} {\bibinfo
  {journal} {Journal of Physics B: Atomic, Molecular and Optical Physics}\
  }\textbf {\bibinfo {volume} {35}},\ \bibinfo {pages} {3599} (\bibinfo {year}
  {2002})}\BibitemShut {NoStop}%
\bibitem [{\citenamefont {Leggett}(2001)}]{leggett_bose-einstein_2001}%
  \BibitemOpen
  \bibfield  {author} {\bibinfo {author} {\bibfnamefont {A.~J.}\ \bibnamefont
  {Leggett}},\ }\bibfield  {title} {\emph {\bibinfo {title} {{Bose-Einstein}
  condensation in the alkali gases: Some fundamental concepts}},\ }\href
  {\doibase 10.1103/RevModPhys.73.307} {\bibfield  {journal} {\bibinfo
  {journal} {Reviews of Modern Physics}\ }\textbf {\bibinfo {volume} {73}},\
  \bibinfo {pages} {307} (\bibinfo {year} {2001})}\BibitemShut {NoStop}%
\bibitem [{\citenamefont {Leggett}(2003)}]{leggett_relation_2003}%
  \BibitemOpen
  \bibfield  {author} {\bibinfo {author} {\bibfnamefont {A.~J.}\ \bibnamefont
  {Leggett}},\ }\bibfield  {title} {\emph {\bibinfo {title} {The relation
  between the {Gross-Pitaevskii} and {Bogoliubov} descriptions of a dilute
  {Bose} gas}},\ }\href {\doibase 10.1088/1367-2630/5/1/103} {\bibfield
  {journal} {\bibinfo  {journal} {New Journal of Physics}\ }\textbf {\bibinfo
  {volume} {5}},\ \bibinfo {pages} {103} (\bibinfo {year} {2003})}\BibitemShut
  {NoStop}%
\bibitem [{\citenamefont {Dziarmaga}\ and\ \citenamefont
  {Sacha}(2003)}]{dziarmaga_bogoliubov_2003}%
  \BibitemOpen
  \bibfield  {author} {\bibinfo {author} {\bibfnamefont {J.}~\bibnamefont
  {Dziarmaga}}\ and\ \bibinfo {author} {\bibfnamefont {K.}~\bibnamefont
  {Sacha}},\ }\bibfield  {title} {\emph {\bibinfo {title} {Bogoliubov theory of
  a {Bose}-{Einstein} condensate in the particle representation}},\ }\href
  {\doibase 10.1103/PhysRevA.67.033608} {\bibfield  {journal} {\bibinfo
  {journal} {Physical Review A}\ }\textbf {\bibinfo {volume} {67}},\ \bibinfo
  {pages} {033608} (\bibinfo {year} {2003})}\BibitemShut {NoStop}%
\bibitem [{\citenamefont {Dziarmaga}\ and\ \citenamefont
  {Sacha}(2006)}]{dziarmaga_images_2006}%
  \BibitemOpen
  \bibfield  {author} {\bibinfo {author} {\bibfnamefont {J.}~\bibnamefont
  {Dziarmaga}}\ and\ \bibinfo {author} {\bibfnamefont {K.}~\bibnamefont
  {Sacha}},\ }\bibfield  {title} {\emph {\bibinfo {title} {Images of a
  {Bose}-{Einstein} condensate: diagonal dynamical {Bogoliubov} vacuum}},\
  }\href {\doibase 10.1088/0953-4075/39/1/007} {\bibfield  {journal} {\bibinfo
  {journal} {Journal of Physics B: Atomic, Molecular and Optical Physics}\
  }\textbf {\bibinfo {volume} {39}},\ \bibinfo {pages} {57} (\bibinfo {year}
  {2006})}\BibitemShut {NoStop}%
\bibitem [{\citenamefont {Jiang}\ \emph {et~al.}()\citenamefont {Jiang},
  \citenamefont {Tacla},\ and\ \citenamefont {Caves}}]{jiang_bosonic_2016}%
  \BibitemOpen
  \bibfield  {author} {\bibinfo {author} {\bibfnamefont {Z.}~\bibnamefont
  {Jiang}}, \bibinfo {author} {\bibfnamefont {A.~B.}\ \bibnamefont {Tacla}}, \
  and\ \bibinfo {author} {\bibfnamefont {C.~M.}\ \bibnamefont {Caves}},\
  }\bibfield  {title} {\emph {\bibinfo {title} {Bosonic particle-correlated
  states: A nonperturbative treatment beyond mean field}},\ }\href@noop {}
  {\bibinfo  {journal} {in preparation}\ }\BibitemShut {NoStop}%
\bibitem [{\citenamefont {Villain}\ \emph {et~al.}(1997)\citenamefont
  {Villain}, \citenamefont {Lewenstein}, \citenamefont {Dum}, \citenamefont
  {Castin}, \citenamefont {You}, \citenamefont {Imamo\=glu},\ and\
  \citenamefont {Kennedy}}]{villain_quantum_1997}%
  \BibitemOpen
\bibfield  {journal} {  }\bibfield  {author} {\bibinfo {author} {\bibfnamefont
  {P.}~\bibnamefont {Villain}}, \bibinfo {author} {\bibfnamefont
  {M.}~\bibnamefont {Lewenstein}}, \bibinfo {author} {\bibfnamefont
  {R.}~\bibnamefont {Dum}}, \bibinfo {author} {\bibfnamefont {Y.}~\bibnamefont
  {Castin}}, \bibinfo {author} {\bibfnamefont {L.}~\bibnamefont {You}},
  \bibinfo {author} {\bibfnamefont {A.}~\bibnamefont {Imamo\=glu}}, \ and\
  \bibinfo {author} {\bibfnamefont {T.~A.~B.}\ \bibnamefont {Kennedy}},\
  }\bibfield  {title} {\emph {\bibinfo {title} {Quantum dynamics of the phase
  of a {Bose-Einstein} condensate}},\ }\href {\doibase
  10.1080/09500349708231846} {\bibfield  {journal} {\bibinfo  {journal}
  {Journal of Modern Optics}\ }\textbf {\bibinfo {volume} {44}},\ \bibinfo
  {pages} {1775} (\bibinfo {year} {1997})}\BibitemShut {NoStop}%
\bibitem [{\citenamefont {Danshita}\ \emph {et~al.}(2005)\citenamefont
  {Danshita}, \citenamefont {Egawa}, \citenamefont {Yokoshi},\ and\
  \citenamefont {Kurihara}}]{danshita_collective_2005}%
  \BibitemOpen
  \bibfield  {author} {\bibinfo {author} {\bibfnamefont {I.}~\bibnamefont
  {Danshita}}, \bibinfo {author} {\bibfnamefont {K.}~\bibnamefont {Egawa}},
  \bibinfo {author} {\bibfnamefont {N.}~\bibnamefont {Yokoshi}}, \ and\
  \bibinfo {author} {\bibfnamefont {S.}~\bibnamefont {Kurihara}},\ }\bibfield
  {title} {\emph {\bibinfo {title} {Collective excitations of {Bose-Einstein}
  condensates in a double-well potential}},\ }\href {\doibase
  10.1143/JPSJ.74.3179} {\bibfield  {journal} {\bibinfo  {journal} {Journal of
  the Physical Society of Japan}\ }\textbf {\bibinfo {volume} {74}},\ \bibinfo
  {pages} {3179} (\bibinfo {year} {2005})}\BibitemShut {NoStop}%
\bibitem [{\citenamefont {Trimborn}\ \emph {et~al.}(2008)\citenamefont
  {Trimborn}, \citenamefont {Witthaut},\ and\ \citenamefont
  {Korsch}}]{trimborn_exact_2008}%
  \BibitemOpen
  \bibfield  {author} {\bibinfo {author} {\bibfnamefont {F.}~\bibnamefont
  {Trimborn}}, \bibinfo {author} {\bibfnamefont {D.}~\bibnamefont {Witthaut}},
  \ and\ \bibinfo {author} {\bibfnamefont {H.~J.}\ \bibnamefont {Korsch}},\
  }\bibfield  {title} {\emph {\bibinfo {title} {Exact number-conserving
  phase-space dynamics of the {$M$}-site {Bose-Hubbard} model}},\ }\href
  {\doibase 10.1103/PhysRevA.77.043631} {\bibfield  {journal} {\bibinfo
  {journal} {Physical Review A}\ }\textbf {\bibinfo {volume} {77}},\ \bibinfo
  {pages} {043631} (\bibinfo {year} {2008})}\BibitemShut {NoStop}%
\bibitem [{\citenamefont {Trimborn}\ \emph {et~al.}(2009)\citenamefont
  {Trimborn}, \citenamefont {Witthaut},\ and\ \citenamefont
  {Korsch}}]{trimborn_beyond_2009}%
  \BibitemOpen
  \bibfield  {author} {\bibinfo {author} {\bibfnamefont {F.}~\bibnamefont
  {Trimborn}}, \bibinfo {author} {\bibfnamefont {D.}~\bibnamefont {Witthaut}},
  \ and\ \bibinfo {author} {\bibfnamefont {H.~J.}\ \bibnamefont {Korsch}},\
  }\bibfield  {title} {\emph {\bibinfo {title} {Beyond mean-field dynamics of
  small {Bose-Hubbard} systems based on the number-conserving phase-space
  approach}},\ }\href {\doibase 10.1103/PhysRevA.79.013608} {\bibfield
  {journal} {\bibinfo  {journal} {Physical Review A}\ }\textbf {\bibinfo
  {volume} {79}},\ \bibinfo {pages} {013608} (\bibinfo {year}
  {2009})}\BibitemShut {NoStop}%
\bibitem [{\citenamefont {Ole\'s}\ and\ \citenamefont
  {Sacha}(2008)}]{oles_n-conserving_2008}%
  \BibitemOpen
  \bibfield  {author} {\bibinfo {author} {\bibfnamefont {B.}~\bibnamefont
  {Ole\'s}}\ and\ \bibinfo {author} {\bibfnamefont {K.}~\bibnamefont {Sacha}},\
  }\bibfield  {title} {\emph {\bibinfo {title} {\mbox{$N$}-conserving
  {Bogoliubov} vacuum of a two-component {Bose-Einstein} condensate: Density
  fluctuations close to a phase-separation condition}},\ }\href {\doibase
  10.1088/1751-8113/41/14/145005} {\bibfield  {journal} {\bibinfo  {journal}
  {Journal of Physics A: Mathematical and Theoretical}\ }\textbf {\bibinfo
  {volume} {41}},\ \bibinfo {pages} {145005} (\bibinfo {year}
  {2008})}\BibitemShut {NoStop}%
\bibitem [{\citenamefont {Schachenmayer}\ \emph {et~al.}(2011)\citenamefont
  {Schachenmayer}, \citenamefont {Daley},\ and\ \citenamefont
  {Zoller}}]{schachenmayer_atomic_2011}%
  \BibitemOpen
  \bibfield  {author} {\bibinfo {author} {\bibfnamefont {J.}~\bibnamefont
  {Schachenmayer}}, \bibinfo {author} {\bibfnamefont {A.~J.}\ \bibnamefont
  {Daley}}, \ and\ \bibinfo {author} {\bibfnamefont {P.}~\bibnamefont
  {Zoller}},\ }\bibfield  {title} {\emph {\bibinfo {title} {Atomic matter-wave
  revivals with definite atom number in an optical lattice}},\ }\href {\doibase
  10.1103/PhysRevA.83.043614} {\bibfield  {journal} {\bibinfo  {journal}
  {Physical Review A}\ }\textbf {\bibinfo {volume} {83}},\ \bibinfo {pages}
  {043614} (\bibinfo {year} {2011})}\BibitemShut {NoStop}%
\bibitem [{\citenamefont {Billam}\ \emph {et~al.}(2013)\citenamefont {Billam},
  \citenamefont {Mason},\ and\ \citenamefont
  {Gardiner}}]{billam_second-order_2013}%
  \BibitemOpen
  \bibfield  {author} {\bibinfo {author} {\bibfnamefont {T.~P.}\ \bibnamefont
  {Billam}}, \bibinfo {author} {\bibfnamefont {P.}~\bibnamefont {Mason}}, \
  and\ \bibinfo {author} {\bibfnamefont {S.~A.}\ \bibnamefont {Gardiner}},\
  }\bibfield  {title} {\emph {\bibinfo {title} {Second-order number-conserving
  description of nonequilibrium dynamics in finite-temperature {Bose-Einstein}
  condensates}},\ }\href {\doibase 10.1103/PhysRevA.87.033628} {\bibfield
  {journal} {\bibinfo  {journal} {Physical Review A}\ }\textbf {\bibinfo
  {volume} {87}},\ \bibinfo {pages} {033628} (\bibinfo {year}
  {2013})}\BibitemShut {NoStop}%
\bibitem [{\citenamefont {Jiang}(2014)}]{JiangPhD}%
  \BibitemOpen
  \bibfield  {author} {\bibinfo {author} {\bibfnamefont {Z.}~\bibnamefont
  {Jiang}},\ }\emph {\bibinfo {title} {Particle Correlations in {Bose-Einstein}
  Condensates}},\ \href {http://hdl.handle.net/1928/24564} {Ph.D. thesis},\
  \bibinfo  {school} {University of New Mexico} (\bibinfo {year}
  {2014})\BibitemShut {NoStop}%
\bibitem [{\citenamefont {Pitaevskii}\ and\ \citenamefont
  {Stringari}(2003)}]{pitaevskii_bose-einstein_2003}%
  \BibitemOpen
  \bibfield  {author} {\bibinfo {author} {\bibfnamefont {L.}~\bibnamefont
  {Pitaevskii}}\ and\ \bibinfo {author} {\bibfnamefont {S.}~\bibnamefont
  {Stringari}},\ }\href@noop {} {\emph {\bibinfo {title} {Bose-Einstein
  Condensation}}}\ (\bibinfo  {publisher} {Oxford University Press},\ \bibinfo
  {address} {Oxford},\ \bibinfo {year} {2003})\BibitemShut {NoStop}%
\bibitem [{\citenamefont {Lewenstein}\ and\ \citenamefont
  {You}(1996)}]{lewenstein_quantum_1996}%
  \BibitemOpen
  \bibfield  {author} {\bibinfo {author} {\bibfnamefont {M.}~\bibnamefont
  {Lewenstein}}\ and\ \bibinfo {author} {\bibfnamefont {L.}~\bibnamefont
  {You}},\ }\bibfield  {title} {\emph {\bibinfo {title} {Quantum phase
  diffusion of a {B}ose-{E}instein condensate}},\ }\href {\doibase
  10.1103/PhysRevLett.77.3489} {\bibfield  {journal} {\bibinfo  {journal}
  {Physical Review Letters}\ }\textbf {\bibinfo {volume} {77}},\ \bibinfo
  {pages} {3489} (\bibinfo {year} {1996})}\BibitemShut {NoStop}%
\bibitem [{\citenamefont {Est\'eve}\ \emph {et~al.}(2008)\citenamefont
  {Est\'eve}, \citenamefont {Gross}, \citenamefont {Weller}, \citenamefont
  {Giovanazzi},\ and\ \citenamefont {Oberthaler}}]{esteve_squeezing_2008}%
  \BibitemOpen
  \bibfield  {author} {\bibinfo {author} {\bibfnamefont {J.}~\bibnamefont
  {Est\'eve}}, \bibinfo {author} {\bibfnamefont {C.}~\bibnamefont {Gross}},
  \bibinfo {author} {\bibfnamefont {A.}~\bibnamefont {Weller}}, \bibinfo
  {author} {\bibfnamefont {S.}~\bibnamefont {Giovanazzi}}, \ and\ \bibinfo
  {author} {\bibfnamefont {M.~K.}\ \bibnamefont {Oberthaler}},\ }\bibfield
  {title} {\emph {\bibinfo {title} {Squeezing and entanglement in a
  {Bose-Einstein} condensate}},\ }\href {\doibase 10.1038/nature07332}
  {\bibfield  {journal} {\bibinfo  {journal} {Nature}\ }\textbf {\bibinfo
  {volume} {455}},\ \bibinfo {pages} {1216} (\bibinfo {year}
  {2008})}\BibitemShut {NoStop}%
\bibitem [{\citenamefont {Grond}\ \emph {et~al.}(2009)\citenamefont {Grond},
  \citenamefont {Schmiedmayer},\ and\ \citenamefont
  {Hohenester}}]{grond_optimizing_2009}%
  \BibitemOpen
  \bibfield  {author} {\bibinfo {author} {\bibfnamefont {J.}~\bibnamefont
  {Grond}}, \bibinfo {author} {\bibfnamefont {J.}~\bibnamefont {Schmiedmayer}},
  \ and\ \bibinfo {author} {\bibfnamefont {U.}~\bibnamefont {Hohenester}},\
  }\bibfield  {title} {\emph {\bibinfo {title} {Optimizing number squeezing
  when splitting a mesoscopic condensate}},\ }\href {\doibase
  10.1103/PhysRevA.79.021603} {\bibfield  {journal} {\bibinfo  {journal}
  {Physical Review A}\ }\textbf {\bibinfo {volume} {79}},\ \bibinfo {pages}
  {021603} (\bibinfo {year} {2009})}\BibitemShut {NoStop}%
\bibitem [{\citenamefont {Riedel}\ \emph {et~al.}(2010)\citenamefont {Riedel},
  \citenamefont {B\"ohi}, \citenamefont {Li}, \citenamefont {H\"ansch},
  \citenamefont {Sinatra},\ and\ \citenamefont
  {Treutlein}}]{riedel_atom-chip-based_2010}%
  \BibitemOpen
  \bibfield  {author} {\bibinfo {author} {\bibfnamefont {M.~F.}\ \bibnamefont
  {Riedel}}, \bibinfo {author} {\bibfnamefont {P.}~\bibnamefont {B\"ohi}},
  \bibinfo {author} {\bibfnamefont {Y.}~\bibnamefont {Li}}, \bibinfo {author}
  {\bibfnamefont {T.~W.}\ \bibnamefont {H\"ansch}}, \bibinfo {author}
  {\bibfnamefont {A.}~\bibnamefont {Sinatra}}, \ and\ \bibinfo {author}
  {\bibfnamefont {P.}~\bibnamefont {Treutlein}},\ }\bibfield  {title} {\emph
  {\bibinfo {title} {Atom-chip-based generation of entanglement for quantum
  metrology}},\ }\href {\doibase 10.1038/nature08988} {\bibfield  {journal}
  {\bibinfo  {journal} {Nature}\ }\textbf {\bibinfo {volume} {464}},\ \bibinfo
  {pages} {1170} (\bibinfo {year} {2010})}\BibitemShut {NoStop}%
\bibitem [{\citenamefont {Gross}\ \emph {et~al.}(2010)\citenamefont {Gross},
  \citenamefont {Zibold}, \citenamefont {Nicklas}, \citenamefont {Est\'eve},\
  and\ \citenamefont {Oberthaler}}]{gross_nonlinear_2010}%
  \BibitemOpen
  \bibfield  {author} {\bibinfo {author} {\bibfnamefont {C.}~\bibnamefont
  {Gross}}, \bibinfo {author} {\bibfnamefont {T.}~\bibnamefont {Zibold}},
  \bibinfo {author} {\bibfnamefont {E.}~\bibnamefont {Nicklas}}, \bibinfo
  {author} {\bibfnamefont {J.}~\bibnamefont {Est\'eve}}, \ and\ \bibinfo
  {author} {\bibfnamefont {M.~K.}\ \bibnamefont {Oberthaler}},\ }\bibfield
  {title} {\emph {\bibinfo {title} {Nonlinear atom interferometer surpasses
  classical precision limit}},\ }\href {\doibase 10.1038/nature08919}
  {\bibfield  {journal} {\bibinfo  {journal} {Nature}\ }\textbf {\bibinfo
  {volume} {464}},\ \bibinfo {pages} {1165} (\bibinfo {year}
  {2010})}\BibitemShut {NoStop}%
\bibitem [{\citenamefont {Perelomov}(1977)}]{perelomov_generalized_1977}%
  \BibitemOpen
  \bibfield  {author} {\bibinfo {author} {\bibfnamefont {A.~M.}\ \bibnamefont
  {Perelomov}},\ }\bibfield  {title} {\emph {\bibinfo {title} {Generalized
  coherent states and some of their applications}},\ }\href {\doibase
  10.1070/PU1977v020n09ABEH005459} {\bibfield  {journal} {\bibinfo  {journal}
  {Soviet Physics Uspekhi}\ }\textbf {\bibinfo {volume} {20}},\ \bibinfo
  {pages} {703} (\bibinfo {year} {1977})}\BibitemShut {NoStop}%
\bibitem [{\citenamefont {Hollenhorst}(1979)}]{hollenhorst_quantum_1979}%
  \BibitemOpen
  \bibfield  {author} {\bibinfo {author} {\bibfnamefont {J.~N.}\ \bibnamefont
  {Hollenhorst}},\ }\bibfield  {title} {\emph {\bibinfo {title} {Quantum limits
  on resonant-mass gravitational-radiation detectors}},\ }\href {\doibase
  10.1103/PhysRevD.19.1669} {\bibfield  {journal} {\bibinfo  {journal}
  {Physical Review D}\ }\textbf {\bibinfo {volume} {19}},\ \bibinfo {pages}
  {1669} (\bibinfo {year} {1979})}\BibitemShut {NoStop}%
\bibitem [{\citenamefont {Caves}\ and\ \citenamefont
  {Schumaker}(1985)}]{Caves1985a}%
  \BibitemOpen
  \bibfield  {author} {\bibinfo {author} {\bibfnamefont {C.~M.}\ \bibnamefont
  {Caves}}\ and\ \bibinfo {author} {\bibfnamefont {B.~L.}\ \bibnamefont
  {Schumaker}},\ }\bibfield  {title} {\emph {\bibinfo {title} {New formalism
  for two-photon quantum optics. \uppercase\expandafter{\romannumeral 1}. \
  {Quadrature} phases and squeezed states}},\ }\href {\doibase
  10.1103/PhysRevA.31.3068} {\bibfield  {journal} {\bibinfo  {journal}
  {Physical Review~A}\ }\textbf {\bibinfo {volume} {31}},\ \bibinfo {pages}
  {3068} (\bibinfo {year} {1985})}\BibitemShut {NoStop}%
\bibitem [{\citenamefont {Schumaker}\ and\ \citenamefont
  {Caves}(1985)}]{Schumaker1985a}%
  \BibitemOpen
  \bibfield  {author} {\bibinfo {author} {\bibfnamefont {B.~L.}\ \bibnamefont
  {Schumaker}}\ and\ \bibinfo {author} {\bibfnamefont {C.~M.}\ \bibnamefont
  {Caves}},\ }\bibfield  {title} {\emph {\bibinfo {title} {New formalism for
  two-photon quantum optics. \uppercase\expandafter{\romannumeral 2}. \
  {Mathematical} foundation and compact notation}},\ }\href {\doibase
  10.1103/PhysRevA.31.3093} {\bibfield  {journal} {\bibinfo  {journal}
  {Physical Review~A}\ }\textbf {\bibinfo {volume} {31}},\ \bibinfo {pages}
  {3093} (\bibinfo {year} {1985})}\BibitemShut {NoStop}%
\bibitem [{\citenamefont {Schumaker}(1986)}]{Schumaker1986a}%
  \BibitemOpen
  \bibfield  {author} {\bibinfo {author} {\bibfnamefont {B.~L.}\ \bibnamefont
  {Schumaker}},\ }\bibfield  {title} {\emph {\bibinfo {title} {Quantum
  mechanical pure states with {Gaussian} wave functions}},\ }\href {\doibase
  10.1016/0370-1573(86)90179-1} {\bibfield  {journal} {\bibinfo  {journal}
  {Physics Reports}\ }\textbf {\bibinfo {volume} {135}},\ \bibinfo {pages}
  {317} (\bibinfo {year} {1986})}\BibitemShut {NoStop}%
\bibitem [{\citenamefont {Poulsen}\ and\ \citenamefont
  {M{\o}lmer}(2001)}]{Poulsen2001a}%
  \BibitemOpen
  \bibfield  {author} {\bibinfo {author} {\bibfnamefont {U.~V.}\ \bibnamefont
  {Poulsen}}\ and\ \bibinfo {author} {\bibfnamefont {K.}~\bibnamefont
  {M{\o}lmer}},\ }\bibfield  {title} {\emph {\bibinfo {title} {Positive-$p$
  simulations of spin squeezing in a two-component {Bose} condensate}},\ }\href
  {\doibase 10.1103/PhysRevA.64.013616} {\bibfield  {journal} {\bibinfo
  {journal} {Physical Review~A}\ }\textbf {\bibinfo {volume} {64}},\ \bibinfo
  {pages} {013616} (\bibinfo {year} {2001})}\BibitemShut {NoStop}%
\bibitem [{\citenamefont {Poulsen}(2002)}]{poulsen_bose-einstein_2002}%
  \BibitemOpen
  \bibfield  {author} {\bibinfo {author} {\bibfnamefont {U.~V.}\ \bibnamefont
  {Poulsen}},\ }\emph {\bibinfo {title} {{Bose-Einstein} Condensates:
  Excursions Beyond the Mean Field}},\ \href
  {http://phys.au.dk/fileadmin/site_files/publikationer/phd/Uffe_V_Poulsen.pdf}
  {Ph.D. thesis},\ \bibinfo  {school} {University of Aarhus} (\bibinfo {year}
  {2002})\BibitemShut {NoStop}%
\bibitem [{\citenamefont {Gillet}\ \emph {et~al.}(2014)\citenamefont {Gillet},
  \citenamefont {Garcia-March}, \citenamefont {Busch},\ and\ \citenamefont
  {Sols}}]{gillet_tunneling_2014}%
  \BibitemOpen
  \bibfield  {author} {\bibinfo {author} {\bibfnamefont {J.}~\bibnamefont
  {Gillet}}, \bibinfo {author} {\bibfnamefont {M.~A.}\ \bibnamefont
  {Garcia-March}}, \bibinfo {author} {\bibfnamefont {T.}~\bibnamefont {Busch}},
  \ and\ \bibinfo {author} {\bibfnamefont {F.}~\bibnamefont {Sols}},\
  }\bibfield  {title} {\emph {\bibinfo {title} {Tunneling, self-trapping, and
  manipulation of higher modes of a {Bose-Einstein} condensate in a double
  well}},\ }\href {\doibase 10.1103/PhysRevA.89.023614} {\bibfield  {journal}
  {\bibinfo  {journal} {Physical Review A}\ }\textbf {\bibinfo {volume} {89}},\
  \bibinfo {pages} {023614} (\bibinfo {year} {2014})}\BibitemShut {NoStop}%
\bibitem [{\citenamefont {Kitagawa}\ and\ \citenamefont
  {Ueda}(1993)}]{kitagawa_squeezed_1993}%
  \BibitemOpen
  \bibfield  {author} {\bibinfo {author} {\bibfnamefont {M.}~\bibnamefont
  {Kitagawa}}\ and\ \bibinfo {author} {\bibfnamefont {M.}~\bibnamefont
  {Ueda}},\ }\bibfield  {title} {\emph {\bibinfo {title} {Squeezed spin
  states}},\ }\href {\doibase 10.1103/PhysRevA.47.5138} {\bibfield  {journal}
  {\bibinfo  {journal} {Physical Review A}\ }\textbf {\bibinfo {volume} {47}},\
  \bibinfo {pages} {5138} (\bibinfo {year} {1993})}\BibitemShut {NoStop}%
\bibitem [{\citenamefont {Sinatra}\ and\ \citenamefont
  {Castin}(2000)}]{sinatra_binary_2000}%
  \BibitemOpen
  \bibfield  {author} {\bibinfo {author} {\bibfnamefont {A.}~\bibnamefont
  {Sinatra}}\ and\ \bibinfo {author} {\bibfnamefont {Y.}~\bibnamefont
  {Castin}},\ }\bibfield  {title} {\emph {\bibinfo {title} {Binary mixtures of
  {Bose-Einstein} condensates: Phase dynamics and spatial dynamics}},\ }\href
  {\doibase 10.1007/s100530050042} {\bibfield  {journal} {\bibinfo  {journal}
  {The European Physical Journal D}\ }\textbf {\bibinfo {volume} {8}},\
  \bibinfo {pages} {319} (\bibinfo {year} {2000})}\BibitemShut {NoStop}%
\bibitem [{\citenamefont {S{\o}rensen}\ \emph {et~al.}(2001)\citenamefont
  {S{\o}rensen}, \citenamefont {Duan}, \citenamefont {Cirac},\ and\
  \citenamefont {Zoller}}]{sorensen_many-particle_2001}%
  \BibitemOpen
  \bibfield  {author} {\bibinfo {author} {\bibfnamefont {A.}~\bibnamefont
  {S{\o}rensen}}, \bibinfo {author} {\bibfnamefont {L.-M.}\ \bibnamefont
  {Duan}}, \bibinfo {author} {\bibfnamefont {J.~I.}\ \bibnamefont {Cirac}}, \
  and\ \bibinfo {author} {\bibfnamefont {P.}~\bibnamefont {Zoller}},\
  }\bibfield  {title} {\emph {\bibinfo {title} {Many-particle entanglement with
  {Bose-Einstein} condensates}},\ }\href {\doibase 10.1038/35051038} {\bibfield
   {journal} {\bibinfo  {journal} {Nature}\ }\textbf {\bibinfo {volume}
  {409}},\ \bibinfo {pages} {63} (\bibinfo {year} {2001})}\BibitemShut
  {NoStop}%
\bibitem [{\citenamefont {Li}\ \emph {et~al.}(2009)\citenamefont {Li},
  \citenamefont {Treutlein}, \citenamefont {Reichel},\ and\ \citenamefont
  {Sinatra}}]{li_spin_2009}%
  \BibitemOpen
  \bibfield  {author} {\bibinfo {author} {\bibfnamefont {Y.}~\bibnamefont
  {Li}}, \bibinfo {author} {\bibfnamefont {P.}~\bibnamefont {Treutlein}},
  \bibinfo {author} {\bibfnamefont {J.}~\bibnamefont {Reichel}}, \ and\
  \bibinfo {author} {\bibfnamefont {A.}~\bibnamefont {Sinatra}},\ }\bibfield
  {title} {\emph {\bibinfo {title} {Spin squeezing in a bimodal condensate:
  spatial dynamics and particle losses}},\ }\href {\doibase
  10.1140/epjb/e2008-00472-6} {\bibfield  {journal} {\bibinfo  {journal} {The
  European Physical Journal B}\ }\textbf {\bibinfo {volume} {68}},\ \bibinfo
  {pages} {365} (\bibinfo {year} {2009})}\BibitemShut {NoStop}%
\bibitem [{\citenamefont {Sinatra}\ \emph {et~al.}(2011)\citenamefont
  {Sinatra}, \citenamefont {Witkowska}, \citenamefont {Dornstetter},
  \citenamefont {Li},\ and\ \citenamefont {Castin}}]{sinatra_limit_2011}%
  \BibitemOpen
  \bibfield  {author} {\bibinfo {author} {\bibfnamefont {A.}~\bibnamefont
  {Sinatra}}, \bibinfo {author} {\bibfnamefont {E.}~\bibnamefont {Witkowska}},
  \bibinfo {author} {\bibfnamefont {J.-C.}\ \bibnamefont {Dornstetter}},
  \bibinfo {author} {\bibfnamefont {Y.}~\bibnamefont {Li}}, \ and\ \bibinfo
  {author} {\bibfnamefont {Y.}~\bibnamefont {Castin}},\ }\bibfield  {title}
  {\emph {\bibinfo {title} {Limit of spin squeezing in finite-temperature
  {Bose-Einstein} condensates}},\ }\href {\doibase
  10.1103/PhysRevLett.107.060404} {\bibfield  {journal} {\bibinfo  {journal}
  {Physical Review Letters}\ }\textbf {\bibinfo {volume} {107}},\ \bibinfo
  {pages} {060404} (\bibinfo {year} {2011})}\BibitemShut {NoStop}%
\end{thebibliography}
\end{document}